\documentclass[9pt,twocolumn,twoside]{pnas-new}

\templatetype{pnasresearcharticle}
\usepackage[utf8]{inputenc}
\usepackage[T1]{fontenc}
\usepackage{graphicx}
\usepackage{amsthm}
\usepackage{amsmath}
\usepackage{amsfonts}
\graphicspath{{./Fig/}}

\graphicspath{{./Fig/}}
\newtheorem*{remark}{Remark}

\def\bv{\bar v}
\def\bn{\bar n}

\title{Plato's cube and the natural geometry of fragmentation}

\author[a,b]{G\'abor Domokos}
\author[c,d,1]{Douglas J. Jerolmack}
\author[e]{Ferenc Kun}
\author[a,f]{J\'anos T\"or\"ok}
\affil[a]{MTA-BME Morphodynamics Research Group, Budapest University of Technology and Economics, Budapest, Hungary}
\affil[b]{Department of Mechanics, Materials and Structure, Budapest University of
Technology and Economics, Budapest, Hungary}
\affil[c]{Department of Earth and Environmental Science,
University of Pennsylvania, Philadelphia, USA}
\affil[d]{Mechanical Engineering and Applied Mechanics,
University of Pennsylvania, Philadelphia, USA}
\affil[e]{Department of Theoretical Physics, University of Debrecen, Debrecen, Hungary}
\affil[f]{Department of Theoretical Physics, Budapest
University of Technology and Economics, Budapest, Hungary}

\leadauthor{G. Domokos}

\significancestatement{We live on and among the byproducts of fragmentation, from nanoparticles to rock falls to glaciers to continents. Understanding and taming fragmentation is central to assessing natural hazards and extracting resources, and even for landing probes safely on other planetary bodies. In this study we draw new inspiration from an unlikely and ancient source: Plato, who proposed that the element Earth is made of cubes because they may be tightly packed together. We demonstrate that this idea is essentially correct: appropriately averaged properties of most natural 3D fragments reproduce the topological cube. We use mechanical and geometric models to explain the ubiquity of Plato's cube in fragmentation, and to uniquely map distinct fragment patterns to their formative stress conditions.}

\authorcontributions{GD originated the concept of the paper, led mathematical and philosophical components, and contributed to field data. DJJ contributed to geophysical interpretation, and led formulation of the manuscript. FK led interpretation of fragmentation mechanics, and contributed to field data. JT performed geometric and DEM computations, and led data analysis. All authors contributed to writing the manuscript.}
\correspondingauthor{\textsuperscript{1}To whom correspondence should be addressed. E-mail: sediment@sas.upenn.edu}

\begin{abstract}
Plato envisioned Earth's building blocks as cubes, a shape rarely found in nature. The solar system is littered, however, with distorted polyhedra --- shards of rock and ice produced by ubiquitous fragmentation. We apply the theory of convex mosaics to show that the average geometry of natural 2D fragments, from mud cracks to Earth's tectonic plates, has two attractors: ``Platonic'' quadrangles and ``Voronoi'' hexagons. 
In 3D the Platonic attractor is dominant: remarkably, the average shape of natural rock fragments is cuboid. When viewed through the lens of convex mosaics, natural fragments are indeed geometric shadows of Plato's forms. Simulations show that generic binary breakup drives all mosaics toward the Platonic attractor, explaining the ubiquity of cuboid averages. Deviations from binary fracture produce more exotic patterns that are genetically linked to the formative stress field. We compute the universal pattern generator establishing this link, for 2D and 3D fragmentation.
\end{abstract}

\dates{This manuscript was compiled on \today}
\doi{\url{www.pnas.org/cgi/doi/10.1073/pnas.XXXXXXXXXX}}

\begin{document}

\maketitle 
\thispagestyle{firststyle}
\ifthenelse{\boolean{shortarticle}}{\ifthenelse{\boolean{singlecolumn}}{\abscontentformatted}{\abscontent}}{}

\section*{Introduction}

\begin{figure}[ht!]
\begin{center}
\includegraphics[width=0.95\columnwidth]{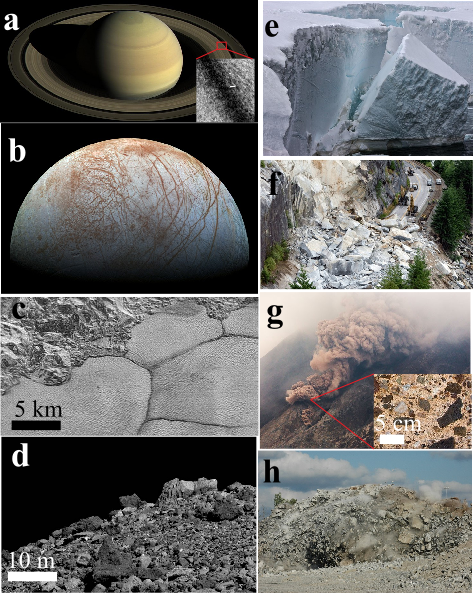}
\caption{\textbf{Fragmentation across planets and scales.} Left column shows planetary surfaces and rings: (a) Saturn's rings composed of ice (inset); (b) Jupiter's moon Europa showing cracked planetary shell; (c) polygonal cracks on Pluto; and (d) surface of the asteroid Bennu. Right column shows example processes forming fragments on Earth: (e) iceberg calving; (f) rock falls; (g) volcanic eruptions that produce pyroclastic flows, forming breccia deposits (inset); and (h) mine blasting. Image credits in Table S2.} 
 
\label{fig:1}
\end{center}
\end{figure}

\dropcap{S}olids are stressed to their breaking point when growing crack networks percolate through the material \cite{turcotte_fractals_1997,kawamura_revmodphys_2012}. Failure by fragmentation may be catastrophic 
\cite{turcotte_fractals_1997,adler_fracture_network_book} (Fig. \ref{fig:1}), but this process is also exploited in industrial applications \cite{prasher1987crushing}. 
Moreover, fragmentation of rock and ice is pervasive within planetary shells \cite{turcotte_fractals_1997,mallard_subduction_2016,mcewen_europa_1986}, and creates granular materials that are literally building blocks for planetary surfaces and rings throughout the solar system \cite{mcewen_europa_1986,trowbridge2016vigorous,cuzzi_saturn_2010,brooker_mars_2018, Brilliantov9536} (Fig. \ref{fig:1}).
Plato postulated that the idealized form of Earth's building blocks is a cube, the only space-filling Platonic solid \cite{domokos_plato_2019, Plato_Timaeus}. 
We now know that there is a zoo of geometrically permissible polyhedra associated with fragmentation \cite{grunbaum1994uniform} (Fig. \ref{fig:2}). Nevertheless, observed distributions of fragment mass \cite{ishii_fragmentation_1992,oddershede_self-organized_1993,wittel_fragmentation_2004,timar_new_2010} and shape \cite{kun_scaling_2006,domokos_universality_2015,ma_shape_2018,szabo_reconstructing_2015} are self-similar, 
and models indicate that geometry (size and dimensionality) matters more than energy input or material composition \cite{steacy_automaton_1991, wittel_fragmentation_2004,astrom_universal_2004} in producing these distributions.  

Fragmentation tiles the Earth's surface with telltale mosaics. Jointing in rock masses forms three-dimensional (3D) mosaics of polyhedra, often revealed to the observer by 2D planes at outcrops (Fig. \ref{fig:2}). The shape and size of these polyhedra may be highly regular, even approaching Plato's cube, or resemble a set of random intersecting planes \cite{dershowitz1988characterizing}. Alternatively, quasi-2D patterns such as columnar joints sometimes form in solidification of volcanic rocks \cite{goering_evolv_pattern_2013}. These patterns have been reproduced in experiments of mud and corn starch cracks, model 2D fragmentation systems, where the following have been observed: fast drying produces strong tension that drives the formation of primary (global) cracks that criss-cross the sample and make ``X'' junctions \cite{ma2019universal,goering_evolv_pattern_2013,desiccation_book_nakahara} (Fig. \ref{fig:3}); slow drying allows the formation of secondary cracks that terminate at ``T'' junctions \cite{ma2019universal}; and ``T'' junctions rearrange into ''Y'' junctions \cite{goering_softmat_2010, goering_evolv_pattern_2013} to either maximize energy release as cracks penetrate the bulk \cite{columnar_joint_science_1988,columnar_joint_jagla_pre2002,hofmann_prl_2015}, or during reopening-healing cycles from wetting/drying \cite{cho2019crack} (Fig. \ref{fig:3}). 
 Whether in rock, ice or soil, the fracture mosaics cut into stressed landscapes (Fig. \ref{fig:3}) form pathways for focused fluid flow, dissolution and erosion that further disintegrate these materials \cite{clair2015geophysical, voigtlander2019breaking} and reorganize landscape patterns \cite{molnar2007tectonics, dibiase2018fracture}. Moreover, fracture patterns in rock determine the initial grain size of sediment supplied to rivers \cite{sklar2017problem, dibiase2018fracture}. 

Experiments and simulations provide anecdotal evidence that the geometry of fracture mosaics is genetically related to the formative stress field \cite{tensilerock_book_2005}. It is difficult to determine, however, if similarities in fracture patterns among different systems are more than skin deep. First, different communities use different metrics to describe fracture mosaics and fragments, inhibiting comparison among systems and scales. Second, we do not know whether different fracture patterns represent distinct universality classes, or are merely descriptive categories applied to a pattern continuum. Third, it is unclear if and how 2D systems map to 3D. 

Here we introduce the mathematical framework of convex mosaics \cite{schneider2008stochastic} to the fragmentation problem. 
This approach relies on two key principles: that fragment shape can be well approximated by convex polytopes \cite{dershowitz1988characterizing}
(2D polygons, 3D polyhedra; Fig. \ref{fig:2}a); and that these shapes must fill space without gaps, since fragments form by the disintegration of solids.
Without loss of generality (SI Section 1.1), we choose a model that ignores the local texture of fracture interfaces \cite{mills1991fractography, bahat2005fractography}.

Fragments can then be regarded as the \emph{cells} of a \emph{convex mosaic} \cite{schneider2008stochastic}, which may be statistically characterized by three parameters. \emph{Cell degree} ($\bar v$) is the average number of vertices of the polytopes, and \emph{nodal degree} ($\bar n$) is the average number of polytopes meeting at one vertex \cite{domokos2019honeycomb}: we call $[\bar n, \bar v] $ the \emph{symbolic plane}. We define the third parameter $0 \leq p \equiv N_R/(N_R+N_I) \leq 1$ as the \emph{regularity} of the mosaic. $N_R$ is the number of \emph{regular} nodes in which cell vertices only coincide with other vertices, corresponding in 2D to ``X'' and ``Y'' junctions with $n=$ 4 and 3, respectively. $N_I$ is the number of \emph{irregular} nodes where vertices lie along edges (2D) or faces (3D) of other cells, corresponding in 2D to ``T'' junctions of $ n = 2$ (Fig. \ref{fig:2}b).  We define a regular (irregular) mosaic as having $p=1$ ($p=0$).
For 3D mosaics we also introduce $\bar f$ as the average number of faces. In contrast to other descriptions of fracture networks \cite{dershowitz1988characterizing}, our framework does not delineate \emph{stochastic} from \emph{deterministic} mosaics; networks made from random or periodic fractures may have identical parameter values (Fig. \ref{fig:3}). 
This theory provides a global chart of geometrically admissible 2D and 3D mosaics in the symbolic plane.  Although we focus here on mosaics formed by fracture, these global charts include all geometrically-possible mosaics, including human-made ones.

In this paper we measure the geometry of a wide variety of natural 2D fracture mosaics and 3D rock fragments, and find that they form clusters within the global chart. Remarkably, the most significant cluster corresponds to the ``Platonic attractor'': fragments with cuboid averages.
Discrete Element Method (DEM) simulations of fracture mechanics show that cuboid averages emerge from primary fracture under the most generic stress field. Geometric simulations show how secondary fragmentation by binary breakup drives any initial mosaic toward cuboid averages.

\section*{2D mosaics in theory and in nature} \label{s:2D}
The geometric theory of 2D convex mosaics is essentially complete \cite{schneider2008stochastic} and is given by the formula \cite{domokos2019honeycomb}: 
\begin{equation}\label{e:2D}
\bar v= \frac{2\bar n}{\bar n - p - 1},
\end{equation}
which delineates the admissible domain for convex mosaics within the $[\bar n, \bar v]$ symbolic plane (Fig. \ref{fig:3}) --- i.e., the global chart. Boundaries on the global chart are given by: (i) the $p=1$ and $p=0$ lines; and (ii) the overall constraints that the minimal degree of regular nodes and cells is 3, while the minimal degree of irregular nodes is 2. We constructed geometric simulations of a range of stochastic and deterministic mosaics (see SI Section 2) to illustrate the continuum of patterns contained within the global chart (Fig. \ref{fig:3})

We describe two important types of mosaics, which help to organize natural 2D patterns.
First are \emph{primitive mosaics}, patterns formed by binary dissection of domains. If the dissection is global we have \emph{regular primitive mosaics} ($p=1$) composed entirely of straight lines which, by definition, bisect the entire sample. These mosaics occupy the point $[\bar n, \bar v]$ = $[4,4]$ in the symbolic plane \cite{schneider2008stochastic}.  In nature, the straight lines appear as primary, global fractures. Next, we consider the situation where the cells of a regular primary mosaic are sequentially bisected locally. Irregular (T-type) nodes are created resulting in a progressive decrease $p \rightarrow 0$ and concomitant decrease $\bar n \rightarrow 2$ toward an \emph{irregular primitive mosaic}. The value $\bar v = 4$, however, is unchanged by this process (Fig. \ref{fig:3}) so in the limit we arrive at $[\bar n, \bar v]$ = $[2,4]$. In nature these local bisections correspond to secondary fracturing \cite{adler_fracture_network_book,tensilerock_book_2005}. Fragments produced from primary vs. secondary fracture are indistinguishable. Further, \emph{any initial mosaic} subject to secondary splitting of cells will, in the limit, produce fragments with $\bar v = 4$ (SI Section 1).
Thus, we expect primitive mosaics associated with the line $\bar v = 4$ in the global chart to be an attractor in 2D fragmentation, as noted by \cite{bohn2005hierarchical}, and we expect the average angle to be a rectangle \cite{ma2019universal} (Fig. \ref{fig:2}). We call this the Platonic attractor. As a useful aside, a planar section of a 3D primitive mosaic (e.g., a rock outcrop) is itself a 2D primitive mosaic (Fig. \ref{fig:2}). The second important pattern is \emph{Voronoi mosaics} which are, in the averaged sense, hexagonal tilings $[\bar n, \bar v]$ = $[3,6]$. They occupy the peak of the 2D global chart (Fig. \ref{fig:3}).

We measured a variety of natural 2D mosaics (SI Section 2) and found, encouragingly, that they all lie within the global chart permitted by Eq. \ref{e:2D}. 
Mosaics close to the Platonic ($\bar v = 4$) line include patterns known  or suspected to arise under primary and/or secondary fracture: jointed rock outcrops, mud cracks, and polygonal frozen ground. 
Mosaics close to Voronoi include mud cracks and, most intriguingly, Earth's tectonic plates. Hexagonal mosaics are known to arise in the limit for systems subject to repeated cycles of fracturing and healing 
\cite{goering_evolv_pattern_2013} (Fig. \ref{fig:3}). We thus consider Voronoi mosaics to be a second important attractor in 2D.  Horizontal sections of columnar joints also belong to this geometric class; however, their evolution is inherently 3D, as we discuss below.

It is known that Earth's tectonic plates meet almost exclusively at ``Y'' junctions; there is debate, however, about whether this ``Tectonic Mosaic'' formed entirely from surface fragmentation, or contains a signature of the structure of mantle dynamics underneath \cite{anderson2002many, bird2003updated, mallard_subduction_2016}. We examine the tectonic plate configuration \cite{bird2003updated} as a 2D convex mosaic, treating the Earth's crust as a thin shell. We find $[\bar n, \bar v]$ = $[3.0, 5.8]$, numbers that are remarkably close to a Voronoi mosaic. Indeed, the slight deviation from $[\bar n, \bar v]$ = $[3,6]$ is because the Earth's surface is a spherical manifold, rather than planar (SI Section 2.3). While this analysis doesn't solve the surface/mantle question, it suggests that the  geometry of the Tectonic Mosaic is compatible with an evolution consisting of 
episodes of brittle fracture and healing. 

The rest of our observed natural 2D mosaics plot between the Platonic and Voronoi attractors (Fig. \ref{fig:3}). We suspect that these
landscape patterns, which include mud cracks and permafrost, either: initially formed as regular primitive mosaics and are in various stages of evolution toward the Voronoi attractor; or were Voronoi mosaics that are evolving via secondary fracture towards the Platonic attractor.  For the case of mosaics in permafrost, however, we acknowledge that mechanisms other than fracture --- such as convection --- could also be at play.

\section*{Extension to 3D mosaics} \label{s:3D}
There is no formula for 3D convex mosaics analogous to the $p=1$ line of Eq. \ref{e:2D} that defines the global chart. There exists a conjecture, however, with a strong mathematical basis \cite{domokos2019honeycomb}; at present this conjecture extends only to regular mosaics. We define the \emph{harmonic degree} as $\bar h = \bar n \bar  v / (\bar n + \bar v)$. The conjecture is that $d < \bar h \leq 2^{d-1}$, where $d$ is system dimension. For 2D mosaics we obtain the known result \cite{schneider2008stochastic, domokos2019honeycomb} $\bar h = 2$, consistent with the $p=1$ substitution in Eq. \ref{e:2D}. In 3D the conjecture is equivalent to $3 < \bar h \leq 4$, predicting that all regular 3D convex mosaics live within a narrow band in the symbolic $[\bar n, \bar v]$ plane \cite{domokos2019honeycomb} (Fig. \ref{fig:4}). Plotting a variety of well-studied periodic and random 3D mosaics (SI Section 3), we confirm that all of them are indeed confined to the predicted 3D global chart (Fig. \ref{fig:4}). 
Unlike the 2D case, we cannot directly measure $\bar n$ in most natural 3D systems. We can, however, measure the polyhedral cells (the fragments): the average numbers $\bar f, \bar v$ of faces and vertices, respectively. Values for $[\bar f, \bar v]$ may be plotted in what we call the \emph{Euler plane}, where the lines bounding the permissible domain 
correspond to \emph{simple} polyhedra (upper) where vertices are adjacent to three edges and three faces, and their dual polyhedra which have triangular faces (lower; Fig. \ref{fig:4}).
Simple polyhedra arise as cells of mosaics in which the intersections are generic --- i.e., at most three planes intersect at one point --- and this does not allow for odd values of $v$.

As in 2D, \emph{3D regular primitive mosaics} are created by intersecting global planes. These mosaics occupy the point  $[\bar n, \bar v]$ = $[8,8]$ 
on the 3D global chart (Fig. \ref{fig:4}). Cells of regular primitive mosaics have cuboid averages  $[\bar f, \bar v]=[6,8]$ \cite{schneider2008stochastic, domokos2019honeycomb}.
This is the Platonic attractor, marked by the $\bar v = 8$ line in the global chart. 
\emph{3D Voronoi mosaics}, similar to their 2D counterparts, are associated with the Voronoi tessellation defined by some random process. If the latter is a Poisson process then we obtain \cite{schneider2008stochastic} $[\bar n, \bar v]=[4,27.07], [\bar f, \bar v]=[15.51,27.07]$ (Fig. \ref{fig:4}). 

\emph{Prismatic mosaics} are created by regarding the 2D pattern as a base, that is extended in the normal direction. 
The prismatic mosaic constructed from a 2D primitive mosaic has cuboid averages, and is therefore statistically equivalent to a 3D primitive mosaic. The prismatic mosaic created from a 2D Voronoi base is what we call a \emph{columnar mosaic}, and it has distinct statistical properties: $[\bar n, \bar v]=[6,12], [\bar f, \bar v]=[8,12]$. 
Thus, the three main natural extensions of the two dominant 2D patterns are 3D primitive, 3D Voronoi, and columnar mosaics.

\begin{figure}[ht]
\begin{center}
\includegraphics[width=0.95\columnwidth]{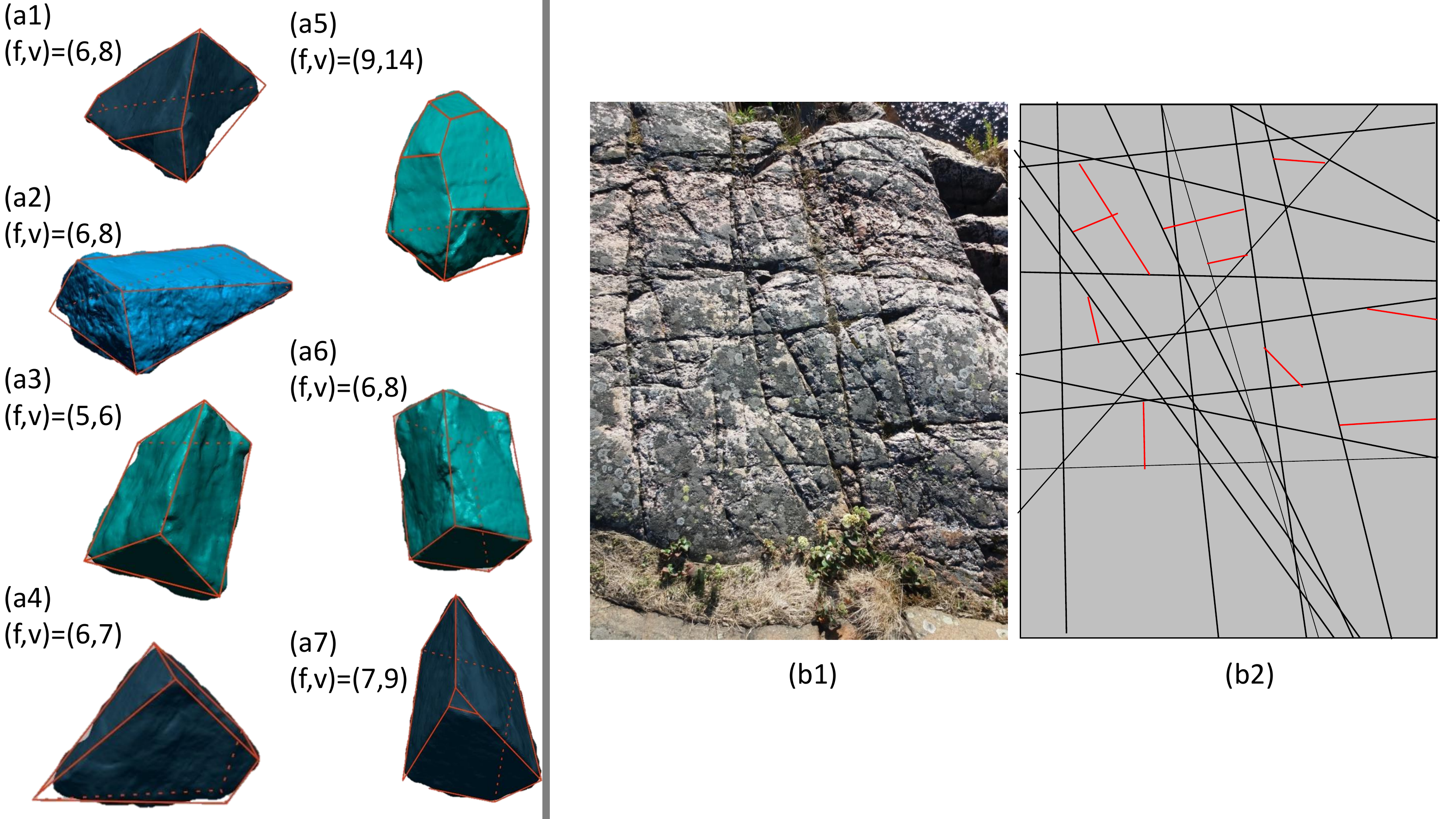}
\caption{\textbf{Examples of fragments and fracture lines.} (a1-a7) Natural fragments approximated by convex polyhedra. (b1) Granite wall showing global cracks. (b2) Approximation of fragmentation pattern by regular
primitive mosaic (black lines) and its irregular version with secondary cracks (red lines). }\label{fig:2}
\end{center}
\end{figure}

\begin{figure}[ht]
\begin{center}
\includegraphics[width=0.95\columnwidth]{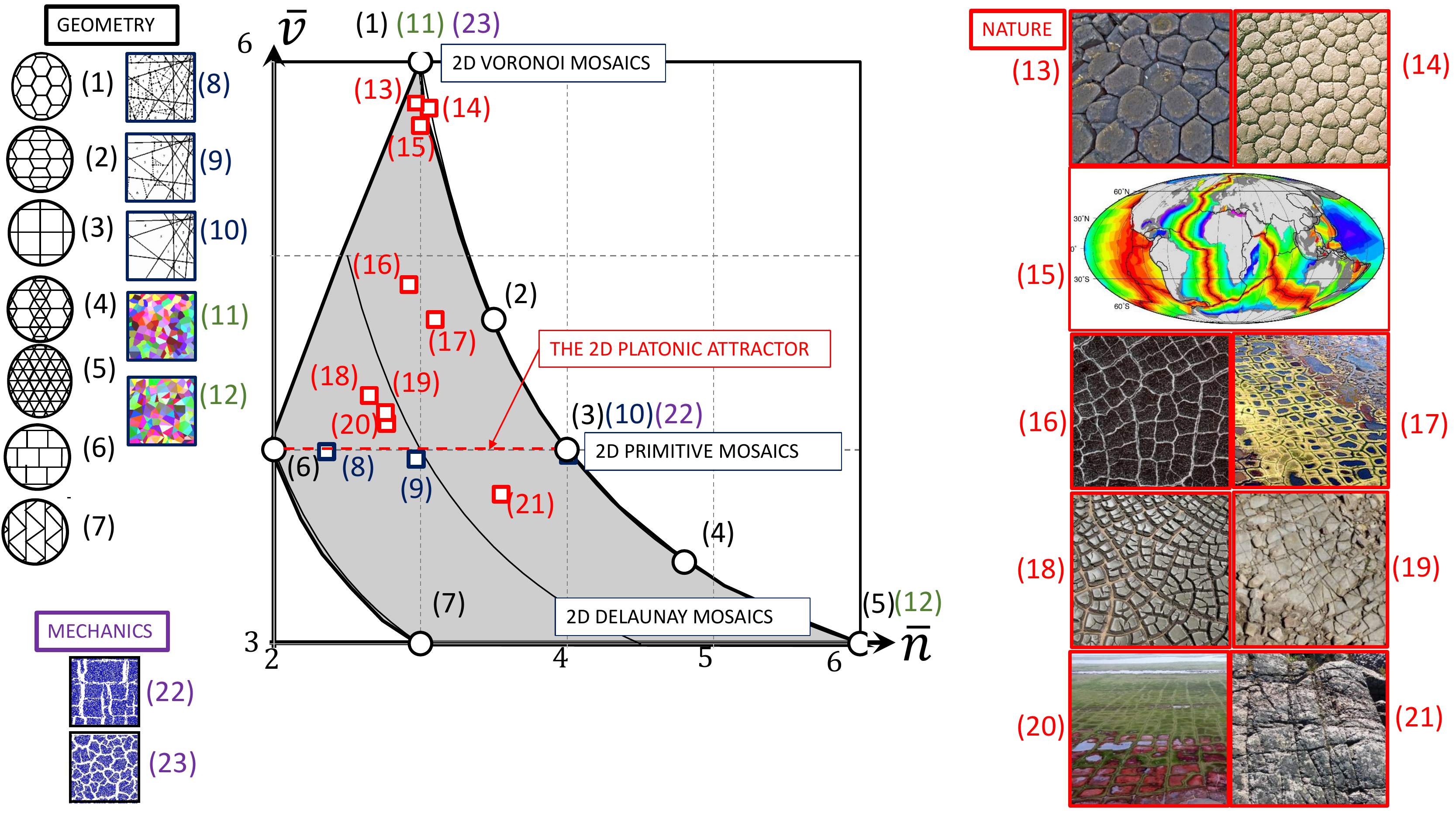}
\caption{\textbf{Mosaics in 2D.} Left: symbolic plane $[\bar n, \bar v]$ with geometrically admissible domain (defined in Eq. \ref{e:2D}) shaded gray.
Patterns (1-7) marked with black circles are deterministic periodic patterns. Patterns (8-12) are geometric simulations of random mosaics: (8) regular
primitive; (9-10) advanced (irregular) primitive; (11)
Poisson-Voronoi; and (12) Poisson--Delaunay.
Red squares (13-21) correspond to analyzed images of natural 2D mosaics shown on the right: (13) columnar joints, Giant's Causeway; (14) mud cracks; (15) tectonic plates; (16) Martian surface; (17) permafrost in Alaska; (18) mud cracks; (19) dolomite outcrop; (20) permafrost in Alaska; and (21) granite rock surface. Image credits in Table S2. Patterns (22-23) are
generated by generic DEM simulation (see Methods): (22) general stress state with eigenvalues $\sigma_1 > \sigma_2$; and (23) isotropic stress state with $\sigma_1 = \sigma_2$. }
\label{fig:3}
\end{center}
\end{figure}

\begin{figure}[ht]
\begin{center}
\includegraphics[width=0.95\columnwidth]{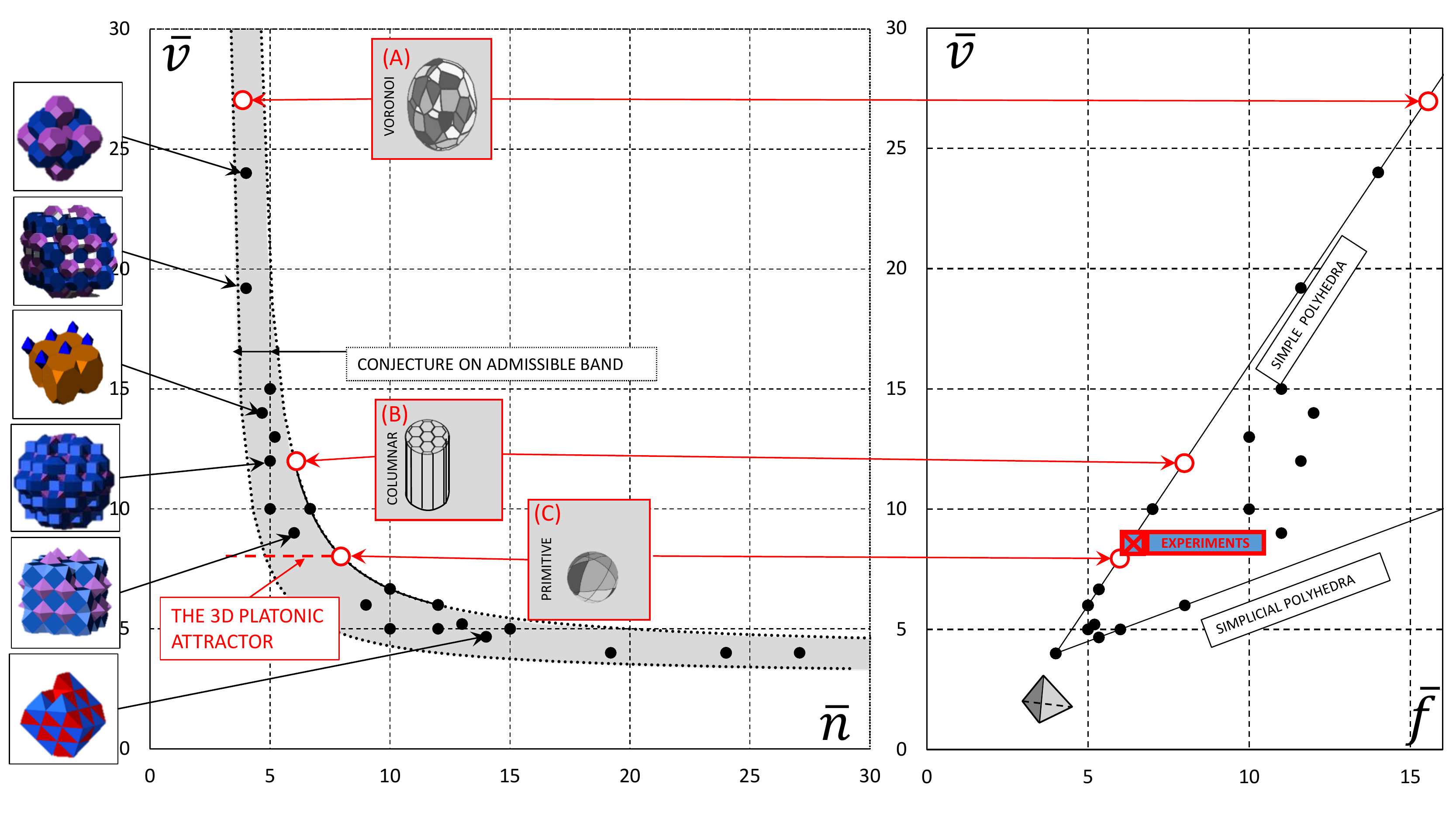}
\caption{\textbf{Mosaics in 3D.} 28 uniform honeycombs, their duals, Poisson-Voronoi, Poisson-Delaunay and primitive random mosaics plotted on the  parameter planes.  Left panel:  $[\bar n, \bar v]$  plane, where continuous black line corresponds to prismatic mosaics \cite{domokos2019honeycomb}. Shaded Gray area marks the predicted domain based on the conjecture $d < \bar h \leq 2^{d-1}$. Right panel:
 $[\bar f, \bar v]$ plane, where straight black lines correspond to simple polyhedra (top) and their duals (bottom); for the tetrahedron $[\bar f, \bar v]= [4,4]$, these two are identical. Mosaics highlighted in Figure \ref{fig:5} are marked by red circles on both panels: (A) 3D Voronoi; (B) columnar mosaics; and (C) 3D primitive mosaic.  }
\label{fig:4}
\end{center}
\end{figure}

Regular primitive mosaics appear to be the dominant 3D pattern resulting from \emph{primary} fracture of brittle materials \cite{hernandez_physa_1995}. Moreover, dynamic brittle fracture produces binary breakup in secondary fragmentation \cite{kekalainen_solution_2007, astrom_universal_2004},
driving the 3D averages $[\bar f, \bar v]$ towards the  Platonic attractor. The most common example in nature is fractured rock (Figs. \ref{fig:2}, \ref{fig:4}). The other two 3D mosaics are more exotic, and seem to require more specialized conditions to form in nature. Columnar joints like the celebrated Giants Causeway, formed by the cooling of large basaltic rock masses \cite{hofmann_prl_2015,lamur_disclosing_2018,goering_evolv_pattern_2013} (Fig. \ref{fig:5}(B)), appear to correspond to columnar mosaics. 
In these systems, the hexagonal arrangement and downward (normal) penetration of cracks arise as a consequence of maximizing energy release \cite{columnar_joint_science_1988,columnar_joint_jagla_pre2002,hofmann_prl_2015}.
The only potential examples of 3D Voronoi mosaics that we know of are septarian nodules, such as the famous Moeraki Boulders \cite{moeraki_boulders_1985} (Fig. \ref{fig:4}). These enigmatic concretions have complex growth and compaction histories, and contain internal cracks that intersect the surface  \cite{astin1986septarian}. Similar to primitive mosaics, the intersection of 3D Voronoi mosaics with a surface is a 2D Voronoi mosaic.

\section*{Connecting primary fracture patterns to mechanics with simulations}\label{s:simulation}

\begin{figure}[ht]
\begin{center}
\includegraphics[width=0.95\columnwidth]{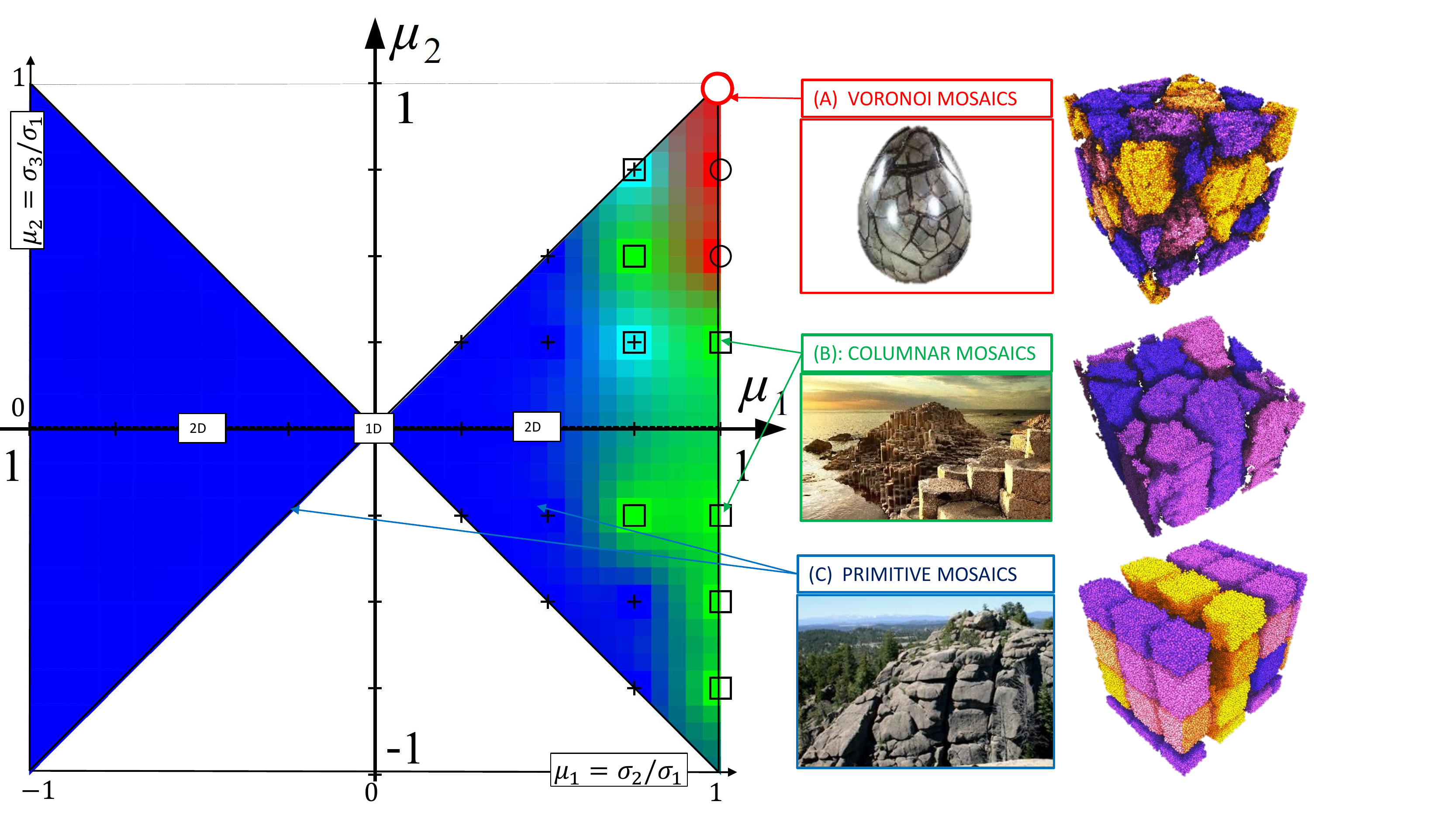}
\caption{\textbf{Illustration of the 3D pattern generator.} Left: the $i=1$ leaf of the $[\mu_1,\mu_2]$ plane. Colors indicate patterns observed in 74 DEM simulations. Blue: primitive, green: columnar, red: Voronoi. Right: (A) 3D Voronoi-type mosaic at $\mu_1=\mu_2=i=1$. Observe surface patterns on all planar sections agreeing with 2D Voronoi-type tessellations, and also with surface patterns of the shown septarian nodule. (B) Columnar mosaic at $\mu_1= 1, \mu_2=0, i=1$. Observe surface pattern on horizontal section agreeing with 2D Voronoi-type tessellation and parallel vertical lines on vertical sections, both matching observation on the illustrated basalt columnar joints. (C) 3D primitive mosaic at $\mu_1=-0.5$, $\mu_2=-0.25$, $i=1$. Observe surface patterns on all sections agreeing with 2D primitive mosaics, and also corresponding to fracture patterns on the illustrated rock. Image credits in Table S2.}
\label{fig:5}
\end{center}
\end{figure}

We hypothesize that primary fracture patterns are genetically linked to distinct stress fields, in order of most generic to most rare. In a 2D homogeneous stress field we may describe the stress tensor with eigenvalues $|\sigma _1|\geq|\sigma _2|$ and characterize the stress state by the dimensionless parameters $\mu = \sigma _2/\sigma _1$ and  $i = \mathrm{sgn}(\sigma _1)$, whose admissible domain is  $\mu \in [-1,1]$  (and it is double covered due to $i = \pm 1$). In 3D this corresponds to eigenvalues $ |\sigma _1|\geq|\sigma _2|\geq |\sigma _3| $, stress state $\mu_1 = \sigma _2/\sigma _1, \mu_2=\sigma_3 / \sigma_1, i = \mathrm{sgn}(\sigma _1)$, and domain $\mu_1, \mu_2 \in [-1,1], |\mu _1| \geq |\mu _2|$ (and it is double covered due to $i = \pm 1$) (Fig. \ref{fig:5}). There is a unique map from these stress-field parameters to the location of the resultant fracture mosaics in the global chart; we call this map the mechanical \emph{pattern generator}. (Results may be equivalently cast on the Flinn diagram \cite{flinn1962diagram} commonly used in structural geology: Fig. S10).
The 2D pattern generator is described by the single scalar function  $\bar v(\mu,i)$  (and $\bar n$ can be computed from Eq. \ref{e:2D}), while the 3D pattern generator is characterized by scalar functions $\bar n(\mu_1,\mu_2, i), \bar v(\mu_1,\mu_2, i), \bar f(\mu_1,\mu_2, i)$. Computing the full pattern generator is beyond the scope of the current paper, even in 2D. Instead, we perform generic DEM simulations \cite{plimpton1995fast,brendel2011contact} of a range of scenarios to interpret the important primary mosaic patterns described above (see Methods).

In 2D, we find that pure shear produces regular primitive mosaics (Fig. \ref{fig:3}), implying $\bar v(-1,\pm 1) \approx 4$. This corresponds to the Platonic attractor. In contrast, hydrostatic tension creates regular Voronoi mosaics (Fig. \ref{fig:3}; SI Section 4) such that  $\bar v(1,1) \approx 6$ --- the Voronoi attractor. Both are in agreement with our expectations.

In 3D, we first 
conducted DEM simulations of hard materials
at $[\mu_1,\mu_2]$ locations corresponding to shear, uniform 2D tension and uniform 3D tension  ($[-0.5,-0.25],[1,-0.2]$ and $[1,1]$, respectively) (Methods, SI Section 4). 
The resulting mosaics displayed the expected fracture patterns for brittle materials: primitive, columnar and Voronoi, respectively (Fig.~\ref{fig:5}). To obtain a global, albeit approximate, picture of the 3D pattern generator we ran additional DEM simulations that uniformly sampled the stress space 
on a $9 \times 9$ ($\Delta \mu{=}0.25$) grid, for both $i{=}+1$ and $i{=}-1$ (SI Section 4). 
The constructed pattern generator demarcates the boundaries in stress-state space that separate the three primary fracture patterns (Fig. \ref{fig:5}). The vast proportion of this space is occupied by primitive mosaics, which are also the only pattern generated under negative volumetric stress. Such compressive stress conditions are pervasive in natural rocks. Columnar 
mosaics are a distant second in terms of frequency of occurrence; they occupy a narrow stripe in the stress space. Most rare are Voronoi mosaics which only occur in a single corner of the stress space (Fig. \ref{fig:5}). 
Boundaries separating the three patterns shifted somewhat for simulations that used softer materials (Fig. S9), but the ranking did not. 
These primary fracture mosaics serve as initial conditions for secondary fracture. While our DEM simulations do not model secondary fracture, we remind the reader that binary breakup drives any initial mosaic toward an irregular primitive mosaic with cuboid averages (SI Section 1) --- emphasizing the strength of the Platonic attractor.

\section*{Geometry of natural 3D fragments}\label{s:data}
\begin{figure}[ht]
\begin{center}
\includegraphics[width=0.95\columnwidth]{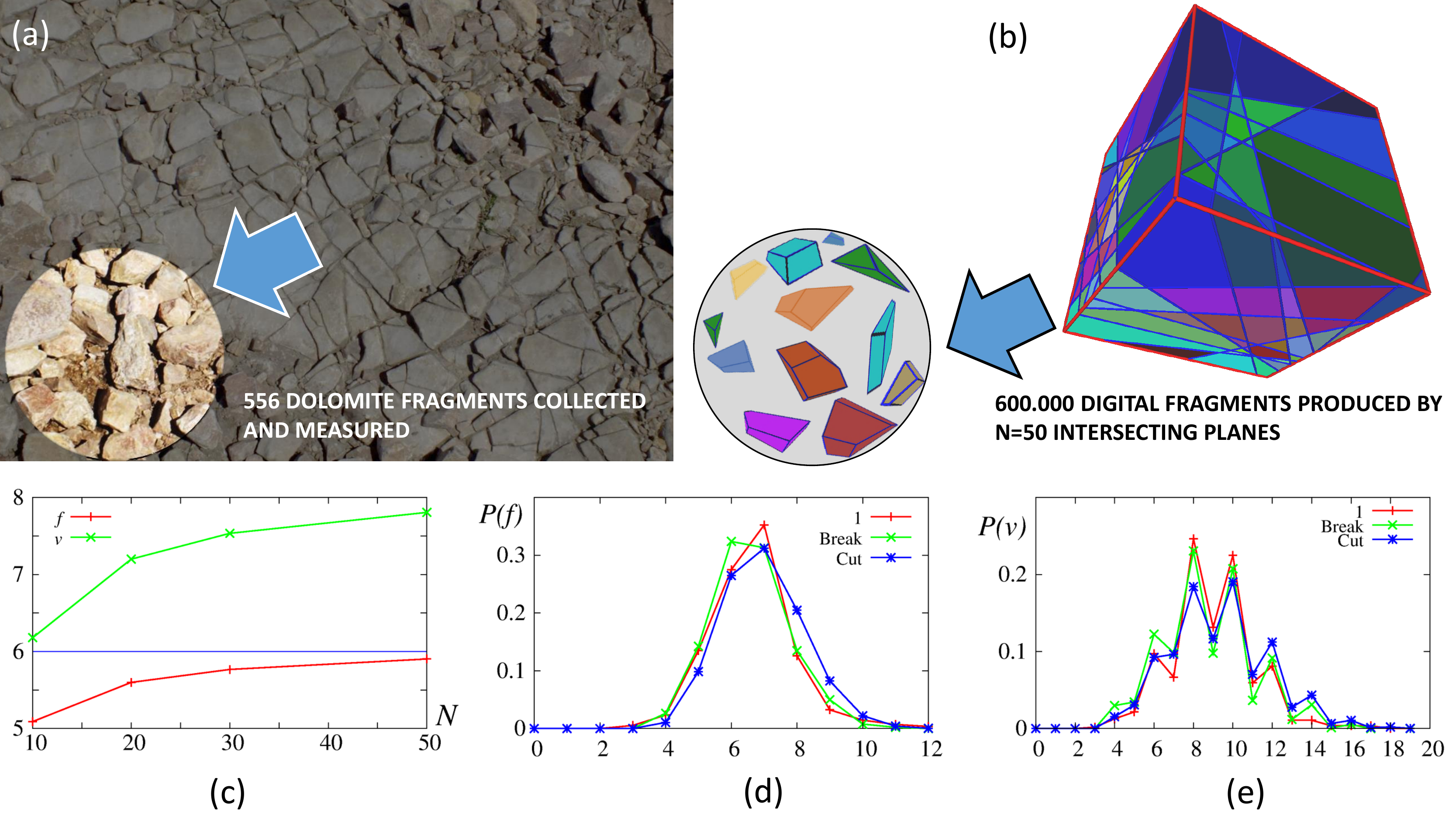}
\caption{\textbf{Natural rock fragments and geometric modeling.} (a): Dolomite rock outcrop at H\'armashat\'arhegy, Hungary, from which we sampled and measured natural fragments accumulating at its base, highlighted in inset. 
(b): The cut model shown with $N= 50$ intersecting planes, and examples of digital fragments (inset) drawn from the 600,000 fragments produced. (c): Evolution of the face and vertex number averages $\bar f$ and $\bar v$ showing convergence toward the cuboid values of 8 and 6, respectively, with increasing $N$. (d) and (e): Probability distributions of $f$ and $v$, respectively, for natural dolomite fragments and fits of the cut and break models (for details on the latter see main text).} 
\label{fig:6}
\end{center}
\end{figure}

Based on the pattern generator (Fig. \ref{fig:5}) we expect that natural 3D fragments should have cuboid properties on average, $[\bar f, \bar v] = [6,8]$. To test this we collected 556 particles from the foot of a weathering dolomite rock outcrop (Fig. \ref{fig:6}) and measured their values of $f$ and $v$, plus mass and additional shape descriptors (see Methods; SI Section 5). We find striking agreement: the measured averages $[\bar f, \bar v]=[6.63, 8.93]$ are within 12 \% of the theoretical prediction, and distributions for $f$ and $v$ are centered around the theoretical values. Moreover, odd values for $v$ are much less frequent than even values, illustrating that natural fragments are well approximated by simple polyhedra (Fig. \ref{fig:6}).
We regard these results as direct confirmation of the hypothesis, while also recognizing significant variability in the natural data.

To better understand the full distributions of fragment shapes, we used geometric simulations of regular and irregular primitive mosaics. The \emph{cut model} simulates regular primitive mosaics as primary fracture patterns by intersecting an initial cube with global planes (Fig.\ref{fig:6}) while the \emph{break model} simulates irregular primitive mosaics resulting from secondary fragmentation processes. We fit both of these models to the shape descriptor data using three parameters: one for the cutoff in the mass distribution, and two accounting for uncertainty in experimental protocols (see Methods; SI Section 5). The best fit model, which corresponds to a moderately irregular primitive mosaic, produced topological shape distributions that are very close to those of natural fragments (mean values $[\bar f, \bar v]=[6.58, 8.74]$).
We also analyzed a much larger, previously collected data set (3728 particles) containing a diversity of materials and formative conditions \cite{domokos_universality_2015}. Although values for $v$ and $f$ were not reported, measured values for classical shape descriptors \cite{domokos_universality_2015,szabo_reconstructing_2015} could be used to fit to the cut and break models (SI Section 5). We find very good agreement ($R^2{>}0.95$), providing further evidence that natural 3D fragments are predominantly formed by binary breakup (Fig. S12). 
Finally, we use the cut model to demonstrate how 3D primitive fracture mosaics converge asymptotically toward the Platonic attractor as more fragments are produced (Fig. \ref{fig:6}).

\section*{Discussion and implications}

The application and extension of the theory of convex mosaics provides a new lens to organize all fracture mosaics --- and the fragments they produce --- into a geometric global chart. There are attractors in this global chart, arising from the mechanics of fragmentation. The Platonic attractor prevails in nature because binary breakup is the most generic fragmentation mechanism, producing averages corresponding to quadrangle cells in 2D and cuboid cells in 3D. Remarkably, a geometric model of random intersecting planes can accurately reproduce the full shape distribution of natural rock fragments. Our findings illustrate the remarkable prescience of Plato's cubic Earth model. One cannot, however, directly `see' Plato's cubes; rather, their shadows are seen in the statistical averages of many fragments. 
The relative rarity of other mosaic patterns in nature make them exceptions that prove the rule. Voronoi mosaics are a second important attractor in 2D systems such as mud cracks, where healing of fractures reorganizes junctions to form hexagonal cells. Such healing is rare in natural 3D systems. Accordingly, columnar mosaics arise only under specific stress fields, that are consistent with iconic basalt columns experiencing contraction under directional cooling. 3D Voronoi mosaics require very special stress conditions, hydrostatic tension, and may describe rare and poorly understood concretions known as septarian nodules.  

We have shown that Earth's Tectonic Mosaic has a geometry that  is compatible with brittle fracture and healing, and is consistent with what is known about plate tectonics  \cite{mallard_subduction_2016} (Fig. \ref{fig:3}). This opens the possibility of inferring stress history from observed fracture mosaics. 
Space missions are accumulating an ever-growing catalogue of 2D and 3D fracture mosaics from diverse planetary bodies, that challenge understanding 
(Fig. \ref{fig:1}). Geometric analysis of surface mosaics may inform debates on planetary dynamics, such as whether Pluto's polygonal surface (Fig. \ref{fig:1}c) is a result of brittle fracture or vigorous convection \cite{trowbridge2016vigorous}. Another potential application is using 2D outcrop exposures to estimate the 3D statistics of joint networks in rock masses, which may enhance prediction of rock fall hazards and fluid flow \cite{national1996rock}. While the present work focused on the shapes of fragments,  the theory of convex mosaics \cite{schneider2008stochastic} is also capable 
of predicting particle-size distributions resulting from fragmentation, which may find application in a wide range of geophysical problems.

The lifecycle of sedimentary particles is a remarkable expression of geometry in nature. Born by fragmentation  \cite{domokos_universality_2015} as statistical shadows of an invisible cube, and rounded during transport along a universal trajectory  \cite{Szabo_universal_2018}, pebble shapes appear to evolve towards the likewise invisible G\"omb\"oc --- albeit without reaching that target \cite{domokos_natural}. The mathematical connections among these idealized shapes, and their reflections in the natural world, are both satisfying and mysterious. Further scrutiny of these connections may yet unlock other surprising insights into nature's shapes.

\section*{Acknowledgements}
The authors thank Zsolt L\'angi for mathematical comments and Kriszti\'an Halmos for his invaluable help with field data measurements. The authors are grateful to David J. Furbish and Mikael Attal whose comments helped to improve the manuscript substantially.
\textbf{Funding:}
This research was supported by: Hungarian NKFIH grants K134199 (G.D.) K116036 (J.T.) and K119967 (F.K.); Hungarian EMMI FIKP grant VIZ (G.D., J.T.); US Army Research Office contract W911-NF-16-1-0290 (D.J.J.); and US National Science Foundation NRI INT award 1734355 (D.J.J.).

\matmethods{
\subsection*{Methods of mechanical simulations}\label{ss:mechanicalmethod}

Initial samples were randomized cubic assemblies of spheres
glued together, with periodic boundary conditions in all directions.The glued contact was realized by a flat elastic
cylinder connecting the two particles which was subject to
deformation from the relative motion of the glued particles. Forces and torques on the particles were calculated based on the deformation of the gluing cylinder.  The connecting cylinder broke permanently if the stress acting upon it exceeded the Tresca criterion \cite{brendel2011contact}. Stress
field was implemented by slowly deforming the underlying space. In order to avoid that there is only one percolating crack, we have set a strong viscous friction between the particles and the underlying space.  The seeding for the evolving cracks was provided by the randomized initial geometry of the spheres.This acts as a homogeneous drag to the particles which ensures a homogeneous stress field in the system.

For any given shear rate the fragment size is controlled by the particle space viscosity and the Tresca criterion limit. We set values that produce reasonable-sized fragments relative to our computational domain, allowing us to characterize the mosaics. Another advantage of the periodic system was that we could avoid any wall
effect that would distort the stress field.
We note here that it is possible to slowly add a shear component to the isotropic tensile shear test and obtain a structure which has average values of $\bn$ and $\bv$ that are between the primitive mosaics and the Voronoi case. Details of mechanical simulations are discussed in SI Section 4.

\subsection*{Methods for fitting geometric model results to field data}\label{ss:geometricmethod}
In the simulation we first computed a regular primitive mosaic by dissecting the unit cube with 50 randomly chosen planes, resulting in $6\times 10^5$ fragments.  We refer to this simulation as the \emph{cut model}. Subsequently we further evolved the mosaic by breaking individual fragments. We implemented a standard model of binary breakup \cite{ishii_fragmentation_1992, domokos_universality_2015} to evolve the cube by secondary fragmentation: at each step of the sequence, fragments either break with a probability $p_b$ into two pieces, or keep their current size until the end of the process with a probability $1-p_b$. The cutting plane is placed in a stochastic manner by taking into account that it is easier to break a fragment in the middle perpendicular to its largest linear extent.
 Inspired by similar computational models \cite{domokos_universality_2015}, we used $p_b$
 dependent on axis ratios (see SI Section 5). This computation, which we call the \emph{break model}, provides an
 approximation to an irregular primitive mosaic; this secondary fragmentation process influenced the nodal degree $\bar n$, but not $[\bar v, \bar f]$.
 
In order to compare numerical results with the experimental data
obtained by manual measurements, we
have to take into account several sampling biases. First, there is always a lower cutoff in size for the experimental
samples. We implemented this in simulations by selecting only fragments
with $m> m_0$, $m_0$ being the cutoff threshold. Second, there is
experimental uncertainty when determining shape descriptors --- especially \emph{marginally stable or unstable} equilibria for the larger dataset (see SI Section 5). We implemented this in the computations by letting the location of the center of mass be a random variable with variation $\sigma_0$ chosen to be small with respect to the smallest diameter of the fragment. We kept only those equilibria  which were found in 95\% of the cases. Third, there is experimental uncertainty in finding very small faces. We implemented  this into the computations by assuming that faces smaller than
$A_0P$  will not be found by
experimenters, where $P$ denotes the smallest projected area of the fragment.
Using the above three parameters we fitted the seven computational histograms to the seven experimental ones by minimizing the largest deviation, and we achieved matches with $R_\mathrm{max}^2 \geq 0.95$ from all histograms (see SI Section 5 for details). Results for the small data set are shown in Fig.~\ref{fig:6}. 
Both the small and the large experimental dataset is freely available at \url{https://osf.io/h2ezc/}, simulation tools for the geometric model are freely available at \url{https://github.com/torokj/Geometric_fragmentation}.
}
\showmatmethods{} 

\onecolumn
\setcounter{table}{0}
        \renewcommand{\thetable}{S\arabic{table}}%
        \setcounter{figure}{0}
        \renewcommand{\thefigure}{S\arabic{figure}}%

\section{Summary of SI Materials}
In this Supplementary Information we provide the following: (i) an in-depth discussion of the mathematical background of convex mosaics, and the emergence of cuboid averages (Section \ref{sec:cuboid}); (ii) information on how we computed the geometry of natural 2D convex mosaics, including the Earth's Tectonic Mosaic (Section \ref{SI2}); (iii) more detail on the generation of synthetic 3D mosaics (Section \ref{SI5}); (iv) detailed information on our discrete element method (DEM) simulations, and their use in creation of the 3D pattern generator linking stress states to fracture mosaic patterns (Section \ref{SI6}); and (v) complete information regarding the geometric simulations of the \textit{cut} and \textit{break} models, and details on the data analysis of $\sim$4000 natural rock fragments and their use in validating the geometric simulations (Section \ref{sec:validate} ).

\section{Cuboid averages}
\label{sec:cuboid}
In this section we provide additional information regarding the shape descriptors used in the paper. We also provide more detail of the mathematics of convex mosaics. Using this information, we present arguments showing that cuboid averages should be expected to be dominant, based on our understanding of the geometry and mechanics of fragmentation.

\subsection{Shape descriptors and definition of cuboid averages}
\label{ss:cuboid_descriptors}

In this paper we  characterize collections of polytopes (natural or artificial) by the average values of certain shape descriptors. If any of these averages agrees exactly with the shape descriptor of the regular cube, we say that the collection has a \emph{cuboid average}. Table \ref{tab:p2} lists all shape descriptors used in this paper (we also use the geological aspect ratios $y_1,y_2$, which are closely related to the ratios $Y_1,Y_2$ in this table). 
\begin{table} [ht] 
{
\begin{center}

\begin{tabular}{| c |r||c|p{9cm}|| c|}
\hline
ID.  & Name & notation &  Definition & Cuboid averages   \\
\hline
\hline
 1 & Faces, Vertices & $f,v$ & Number of faces and vertices of the polyhedron, respectively. & 6,8 \\
\hline
 2 & Angles & $\alpha _i, \beta _i$ & Angles between faces sharing a common edge, and edges sharing common vertex, respectively. & $\pi/2, \pi /2$ \\
\hline

 3 & Aspect ratios & $Y_1\geq Y_2$ & $Y_1=c/a, Y_2=b/a$,  $a\geq b \geq c$ are the sizes of the smallest orthogonal bounding box & 1,1 \\
\hline

4 & Equilibria & $S,U$ & Numbers of stable and unstable static balance points & 6,8 \\

\hline
\end{tabular}
\end{center}
}

\vspace{0.5cm}
\caption{\textbf{Shape descriptors}}
\label{tab:p2}

\end{table}

In our paper we make the claim that the majority of collections of natural fragments display cuboid averages for $f,v$ and for $\alpha_i, \beta _i$. We do \textit{not} claim that axis ratios or static equilibria exhibit cuboid averages, and indeed they do not. Nevertheless, the latter shape descriptors are useful for comparing geometric simulations with experimental data in two respects: (i) they represent an additional test making model comparisons to data more stringent; and (ii) they allow us to analyze previously collected data on rock fragments that measured axis ratios and equilibria, but did not report values for faces, vertices or angles. 
Finally, we note that our modeling approach for complex mosaics implicitly treats fractures as planar, and therefore ignores the varied textures of natural fracture interfaces. Since we only aim to approximate global geometric features of fragmentation patterns, and these textures vary on length-scales much smaller than the size of the fragments, this is a safe assumption. Moreover, the theory of convex mosaics describes topological features (numbers of faces and edges) that are insensitive to such textures, allowing its extension to non-planar fractures.


\subsection{Cuboid averages in primitive mosaics}\label{ss:cuboid_primitive}

Primitive mosaics (also known as \emph{hyperplane mosaics} \cite{schneider2008stochastic}) are an important class of random tessellations.
We call a hyperplane mosaic in $d$-dimensions \emph{generic} if each node is generated by the intersection of $d$ hyperplanes of dimension $d-1$.
Most often they are studied in the context of Poisson processes and the related mosaics are called Poisson-hyperplane mosaics \cite{schneider2008stochastic}. It is well known that these mosaics display cuboid averages in any dimension \cite{schneider2008stochastic}. Most notably, the average values \emph{do not depend} on the chosen distribution, i.e. they not only characterize Poisson-hyperplane mosaics but  in fact any generic hyperplane mosaics. It is also known (and in fact, easy to see) that intersecting a $d$-dimensional primitive (hyperplane) mosaic with a $d-1$-dimensional hyperplane results
in a $d-1$-dimensional hyperplane mosaic. Since the formalism in \cite{schneider2008stochastic} may not be easily accessible, below we present a sketch of an elementary proof to illustrate that cuboid averages do not depend on the distributions and they are indeed a fundamental property of primitive mosaics.
\subsection{Sketch of elementary proof of cuboid averages in primitive mosaics in any dimension}
\begin{proof}
First we observe that the Gauss map establishes a bijection $B_1$ between the external surface normals of any cell and the 
points on the sphere. Next we
regard a node which is a point where the vertices of adjacent cells overlap. Here the Gauss map
also provides a bijection $B_2$ between 
the external surface
normals in the small vicinity of those adjacent vertices and the points on the sphere. The
combination of these two bijections $B_3=B_1 \circ B_2$ 
provides a bijection between the nodes and the cells of the primitive mosaic
guaranteeing that, for a sufficiently large mosaic, we have (approximately) the same number of nodes and cells. This implies that if we take
the average of any quantity over all nodes, the same average will apply to the cells.
\end{proof}
\begin{remark}

The above argument relies on the existence of the bijections $B_1$ and $B_2$.
While the former always exists for convex cells, the latter only exists for primitive mosaics.
\end{remark}

In $d=2$ dimensions, if the primitive  mosaic is generic then at each node we have exactly 4 vertices, so for the \emph{average} vertex number of cells we get $\bar v=4.$ Also, at the node the average
angle between any two edges meeting at a vertex is 90 degrees and this will also
apply to the cells.

In $d=3$ dimensions, if the primitive  mosaic is generic then at each node we have exactly 8 vertices of degree 3, i.e. we have $\bar n=8$.
The bijection guarantees that the averages for cells have the same values, resulting in $\bar v=8, \bar f=6$. Also, at the node the average
angle between any two edges meeting at a vertex is 90 degrees and this will also apply to the cells. 

We also remark that our argument does not provide any predictions for the averages of axis ratios $Y_1,Y_2$ and the averages for the numbers $S,U$ of static equilibrium points.

\subsection{Generating mechanisms: expanding domains and binary breakup}\label{ss:generating}

Our main statement about the equivalence of nodal and cell averages applies to \emph{sufficiently large} mosaics, i.e. to mosaics with sufficiently large number of nodes and cells. Averages can be only defined on a finite 
domain, and some limit process is required to produce a sufficiently large domain. 

One possibility is to regard an infinite mosaic, establish a fixed point, and compute the averages for all nodes and cells which lie inside a ball $\mathcal B$ with radius $R$. In this case the $R \to \infty$ limit can provide sufficiently large domains. This procedure may be applied to any Euclidean mosaic.

\begin{figure}[ht]
\begin{center}
\includegraphics[width=0.9\columnwidth]{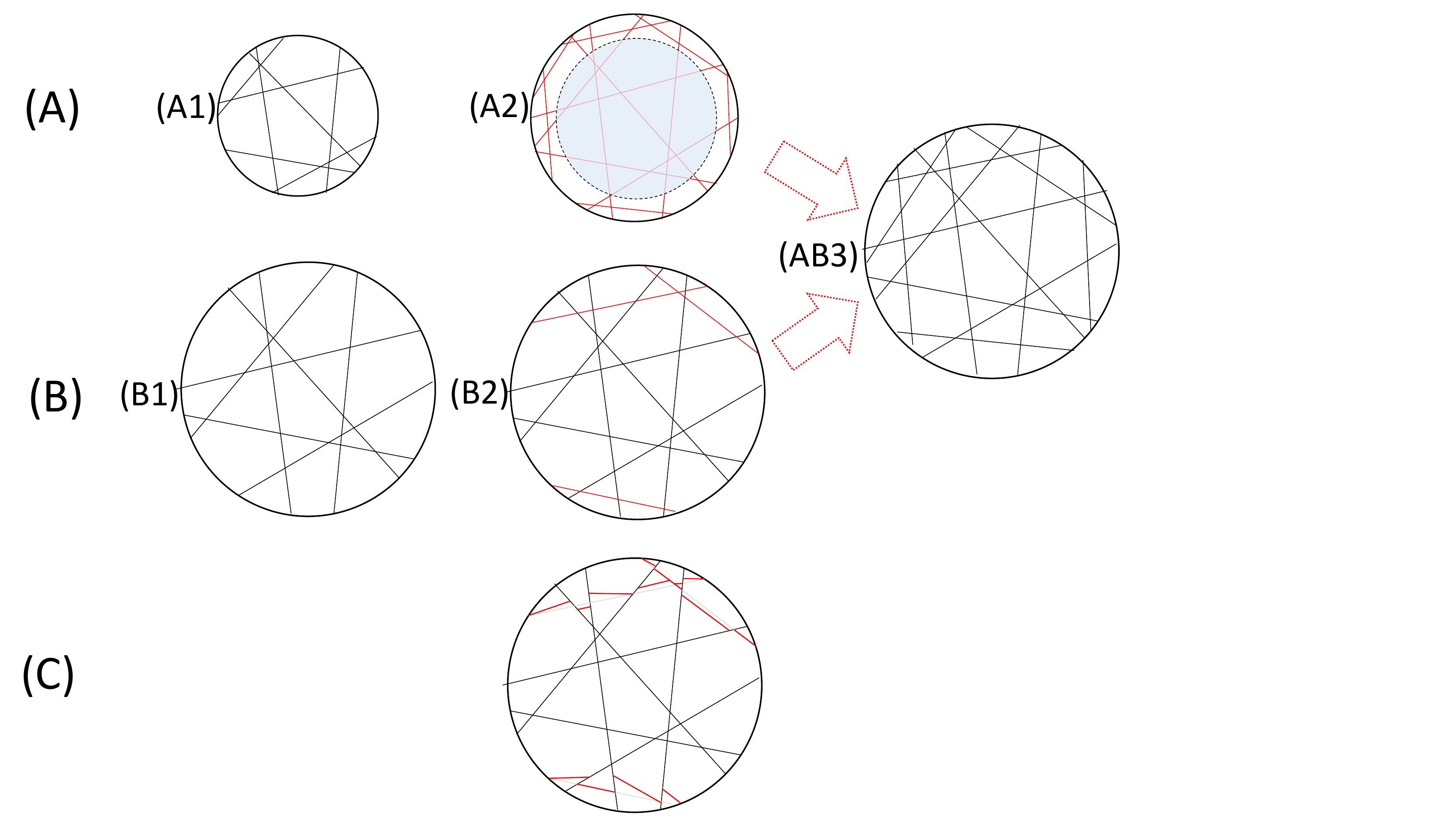}
\caption{\textbf{Alternative limits  and binary  breakup mechanism
characterizing primitive mosaics.} (A) Expanding the domain by increasing the radius on a fixed mosaic. (B) Expanding the domain by adding new lines.  Starting configurations (A1), (B1) evolve via (A2) and (B2) towards (AB3). Red color in (A2) and (B2) indicates the difference with respect to (A1) and (B1), respectively. (C) Configuration obtained from (B1) by individual binary breakup. All cell degrees agree with (B2).}
\label{fig:limits}
\end{center}
\end{figure}

Alternatively, in case of primitive mosaics, one may regard a ball $\mathcal B$ with fixed radius $R$. Assuming that $N$ hyperplanes intersect $\mathcal B$, we may regard the limit $N \to \infty$ as a sufficiently large domain.
Both type of limits are illustrated in Figure \ref{fig:limits}. 

As we can observe, the domain of primitive mosaics is expanded in the second scenario by \emph{binary dissection} of groups of cells. As Figure \ref{fig:limits} (c) illustrates, individual binary breakup (as opposed to breakup defined by a global line) results in a different fragmentation pattern (which we call irregular primitive mosaic)  with different (smaller) nodal degrees. However, we can also observe that in the vicinity of any regular primitive mosaic we have infinitely many irregular primitive mosaics with identical cell averages.

This indicates that individual binary breakup, producing irregular primitive mosaics, results in cell averages identical to regular primitive mosaics. Below we present an elementary argument in $d=2$ dimensions supporting this claim.

\subsection{Sketch of arguments for cuboid averages in 2D and 3D individual binary breakup}
\subsubsection{Sketch of 2D argument}
Consider a convex polygon $P_0$ with $v_0$ vertices and bisect it with  a randomly chosen straight line (the distribution is not relevant for our purpose). Any generic line will intersect two edges of $P_0$ and the resulting two polygons will be $P_{1,1}, P_{1,2}$
with vertex numbers $v_{1,1},v_{1,2}$, respectively.  We define their average as $v_1=(v_{1,1}+v_{1,2})/2$ and we also define the increment $\Delta v_1=v_1-v_0$. We can immediately see that
\begin{equation}
\Delta v_1(v_0-4) \leq 0, \quad \frac{v_1-4}{v_0-4}=\frac{1}{2},
\end{equation}
i.e., $v_1$ will be twice as close to the value 4 as $v_0$ was. This idea can be readily generalized to a hierarchical breakup process
with $k$ subsequent steps, producing $2^k$ fragments with average vertex number $v_k$. From this we can see that 
\begin{equation}
\lim_{k \to \infty}v_k=4.    
\end{equation}
\subsubsection{Sketch of 3D argument}
In 3D we can not offer a mathematical argument completely analogous to the 2D case, i.e. we can not prove that cuboid averages represent an attractor for binary breakup processes, although our numerical simulations (the break model) certainly indicate that this is a fact.  However, as we described in subsection \ref{ss:generating}, the generating mechanism for regular and irregular primitive mosaics are closely related suggesting that the corresponding averages should also be related.

Here we outline a mathematical argument which suggests that if cuboid averages are attractors, then they are likely to be strong attractors.

Halving the volume of a convex polyhedron P with a random cut from
uniformly random direction may serve as a simplistic geometric model
of individual binary breakup. Let us denote the \emph{replicating quality} as the following: the probability that such a halving cut results in two convex polyhedra that are combinatorially equivalent to P.
In 2D the square has replicating quality 100\%, while polygons with V>4
vertices have zero replicating quality. In 3D it is relatively easy to prove that the only canonical polyhedron with replicating quality larger
than 50\% is the cube. In other words, cuboid particles are likely to produce more cuboid particles, while other shapes are unlikely to propagate themselves under bisection.

This model certainly supports the numerical results of the break model, which showed that cuboid averages are a robust attractor of individual binary breakup.

\subsection{Interpretation of cuboid averages}
While cuboid averages strongly suggest that the geometry of the cube  plays a central role in fragmentation processes, one has to be cautious with the interpretation of these results.

\begin{figure}[ht]
\begin{center}
\includegraphics[width=0.8\columnwidth]{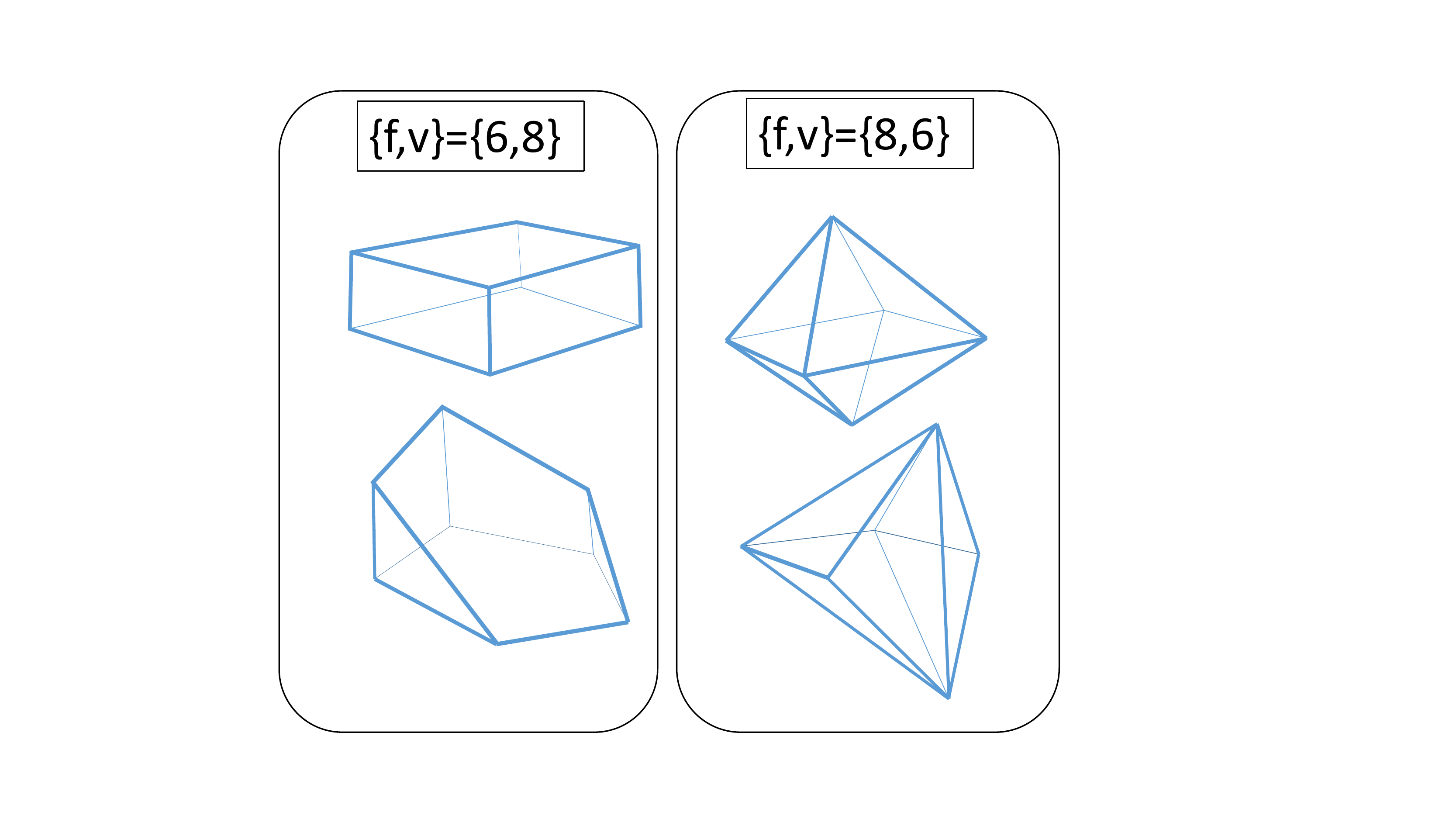}
\caption{\textbf{Secondary class examples.} Secondary, combinatorial classes in the primary classes $\{6,8\}$ and $\{8,6\}.$}
\label{fig:SI_68_86}
\end{center}
\end{figure}

As we stated above, we obtained cuboid averages  $\bar f,\bar v$ for the numbers of faces and vertices. This result does not imply in any manner that the polyhedral approximation of the \emph{the most common} fragment shape should be combinatorially equivalent to the cube. The pair $\{f,v\}$ is often referred to as the \emph{primary class} of a polyhedron which may contain several, combinatorially different \emph{secondary classes}: for a list of combinatorial classes of polyhedra up to 8 faces see \cite{federico1975polygons}. Figure \ref{fig:SI_68_86} illustrates the two pairs of combinatorially different polyhedra appearing in the primary classes $\{ 6,8\}$  and $\{8,6\}.$ 

\begin{figure}[ht!]
\begin{center}
\includegraphics[width=0.7\columnwidth]{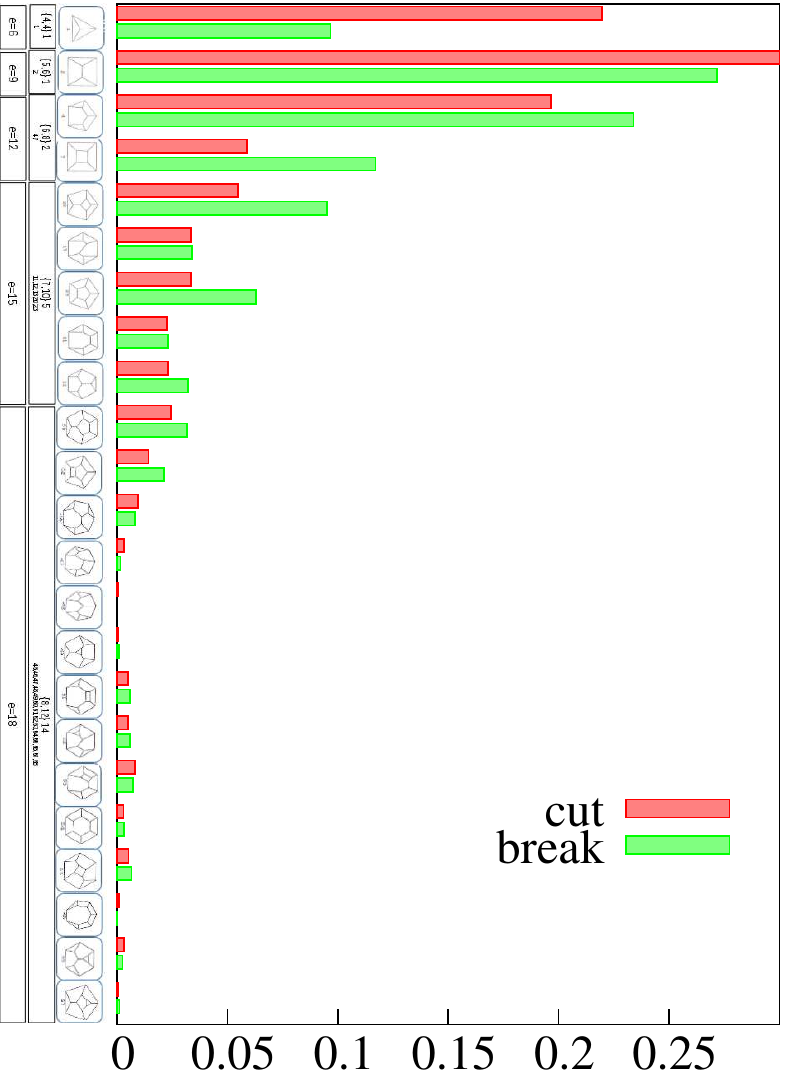}
\caption{\textbf{Statistical distribution of combinatorial classes of
polyhedra in a primitive mosaic.} The columns on the left side indicate the following. First column: number of edges. Second column: the number of faces and edges in curly brackets, the number of subclasses in the same primary class, the code in \protect\cite{federico1975polygons}. Third column: the schematic layout of the polyhedron corresponding to the column of the histogram. Red and green bars belong to the computations based on the cut and break models, respectively. }
\label{fig:SI_combinatorial}
\end{center}
\end{figure}

We do not have any experimental data on combinatorial classes; however, we have a wealth of data from computer simulations. We used this large dataset to compute the statistics for the distribution of combinatorial classes among polyhedra in a primitive mosaic.
We computed the statistics both for the cut model (producing regular primitive mosaics) as well as for the break model (producing irregular primitive mosaics),
both illustrated in Figure \ref{fig:SI_combinatorial}, and we found that the difference between the two models was not substantial in the sense that both models identified the very same combinatorial classes as dominant. In particular, the set of combinatorial classes with relative frequencies above 0.035 agree between the two models. On the other hand, we can observe a marked difference between the break and cut models among tetrahedra which may be related to the average elongation of these fragments. Overall, both (cut and break) histograms appear globally analogous to discretized lognormal distributions. These statistics may also be of broader interest, as there are no theoretical results for these distributions.

\section{2D mosaics}\label{SI2}
Here we discuss details on the generation of synthetic 2D mosaics that are presented in the main text and also provide additional examples. Then we describe the methodology for quantifying the geometry of natural 2D fracture mosaics presented in the main text and we discuss in detail the geometry of the Tectonic Mosaic.

\subsection{2D geometric examples}

\begin{figure}[ht!]
\begin{center}
\includegraphics[width=0.95\columnwidth]{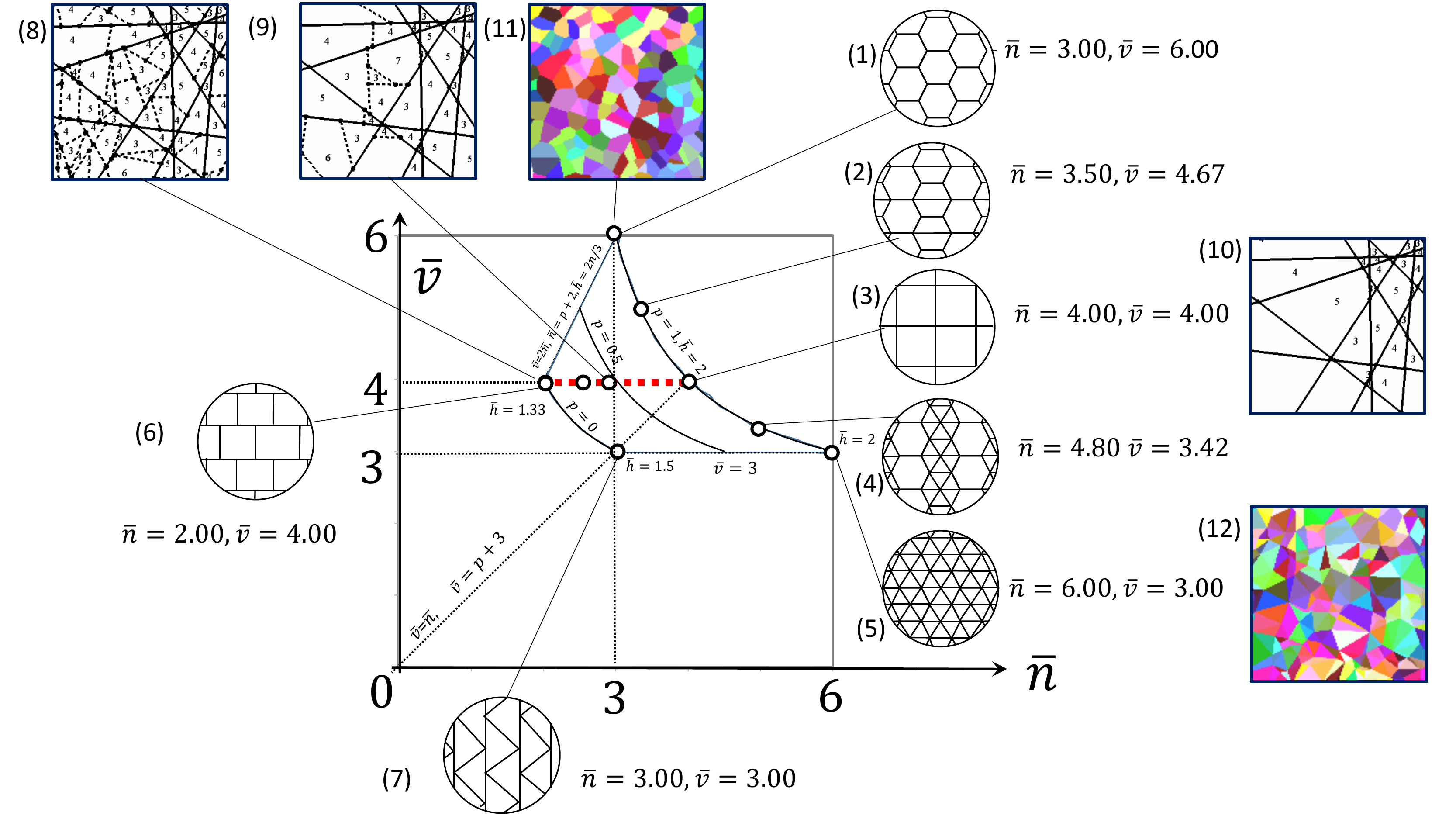}
\caption{\textbf{Geometrical examples for 2D mosaics.}
      (1) Regular, periodic hexagonal lattice. (2) Regular, periodic mixed hexagonal square tiling. (3) Regular, periodic square lattice. (4) Regular, periodic mixed hexagonal triangular tiling. (5) Regular, periodic triangular lattice. (6) Fully irregular, periodic brick tiling. (7) Fully irregular, periodic triangular tiling. (8) Random, almost fully irregular primitive mosaic resulting from the 2D version of the break model with high break probability. (9) Random, moderately irregular primitive mosaic resulting from the 2D version of the break model with low break probability. (10) Random regular primitive mosaic resulting from 2D cut model. (11) Voronoi tessellation of a set of random points. (12) Delaunay triangulation of the same set of random points. Dashed line shows the Platonic attractor for cells, $\bar v = 4$.}
\label{fig:SI_2D:geometry}
\end{center}
\end{figure}

In Figure 3 of the main text we showed seven periodic and five random patterns (numbered consecutively 1-12) as illustrations of planar convex mosaics. In Figure \ref{fig:SI_2D:geometry} we give the
exact $\bar n, \bar v$ values corresponding to these mosaics and provide additional details on the global chart defined by formula (1) of the main text on the symbolic plane $[\bar n, \bar v].$

\subsection{2D natural examples}

Figure 3 of the main text illustrates nine natural examples
(image sources listed in Table \ref{tab:2D_sources}) and indicates the corresponding point on the $[\bar n, \bar v]$ symbolic plane. The values for $\bar n, \bar v$ were obtained in the following manner: the convex mosaic associated with the pictures was extracted by using our own semi-automated image analysis software. Once the mosaic was available, the vertex and nodal degrees were counted and the averages computed. To maintain some control on the consistency of the measurement, we computed the vertex degree $\bar v$ of the mosaic not only as the average of the vertex number of the individual cells, but we also used formula (1) to obtain $\bar v$ from the nodal degree and the regularity parameter $p$.  Next we discuss the analysis of a 2D natural mosaic in detail. 
As an additional example, in Figure \ref{fig:2D_nature} we show the image analysis for the Martian landscape, also shown in Figure 3 of the main text.

\subsubsection{2D analysis of the dolomite outcrop on H\'armashat\'arhegy }

\begin{figure}[ht!]
\begin{center}
\includegraphics[width=0.95\columnwidth]{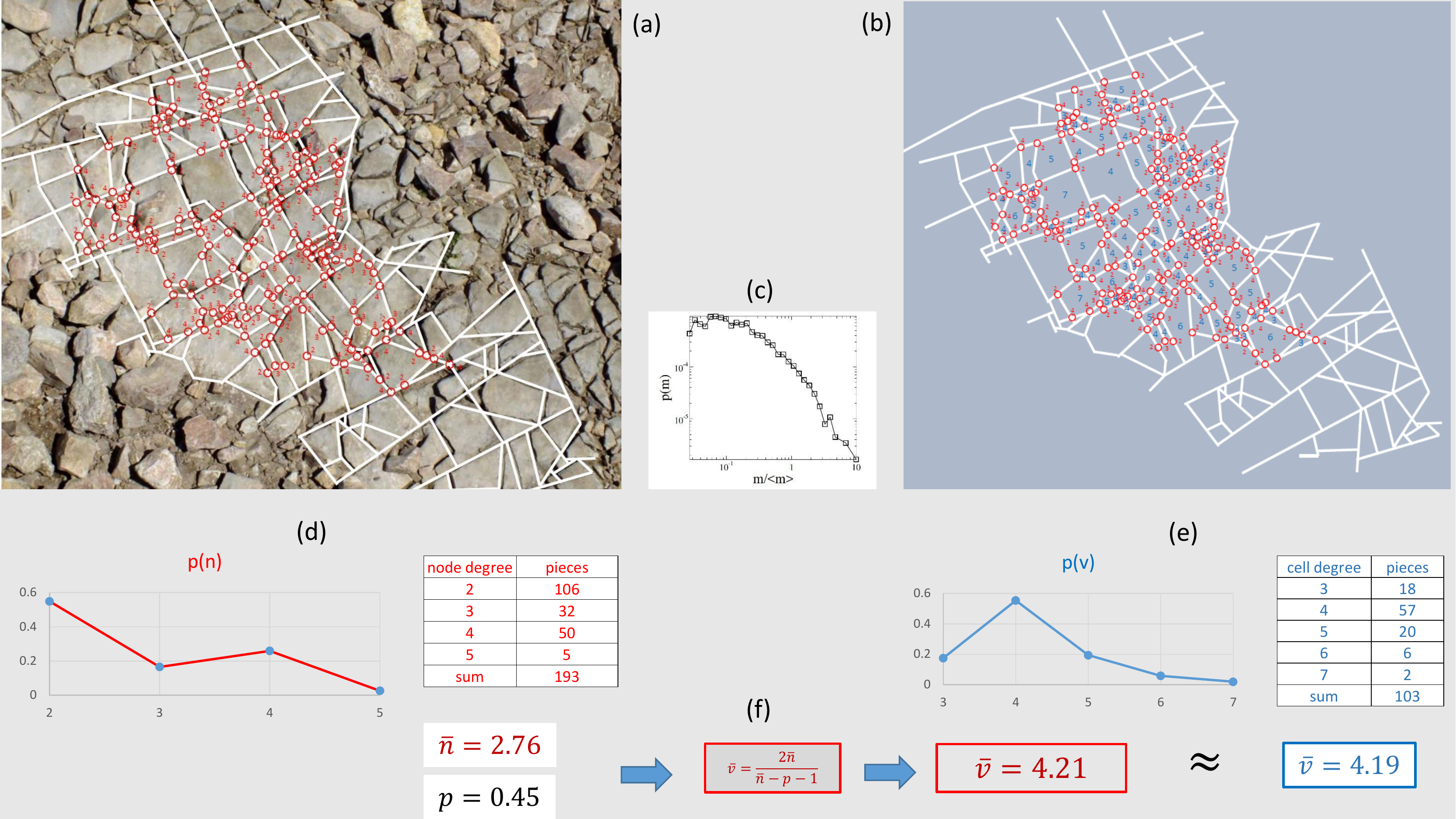}
\caption{\textbf{2D analysis of the dolomite rock outcrop at
H\'armashat\'arhegy.} (a) Image of outcrop with superposed (idealized) fragmentation network. Cracks shown as white lines, nodes marked with red circles. Nodal degrees inserted as red numbers. (b) Fragmentation mosaic without image, with added cell degrees (blue numbers). (c) Mass distribution, where mass was approximated by area. (d) Relative frequencies of nodal degree showing that most nodes were irregular T nodes with degree $n=2$. (e) Relative frequencies for cell degree showing that most cells were quadrangles with $v=4$. (f) Comparison between cell degree $\bar v$ computed by mosaic formula (1) and by direct averaging. Good agreement between the experimentally obtained value and the computed value shows applicability of the formula (originally derived for infinite mosaics) for imperfect (finite) mosaics.
}
\label{fig:2D_Harmashatar}
\end{center}
\end{figure}

As we pointed out in subsection \ref{ss:cuboid_primitive}, lower dimensional sections of primitive  mosaics are themselves primitive mosaics. Since the dataset of 3D fragments obtained from the dolomite outcrop at H\'armashat\'arhegy plays a central role in our argument, we also performed a full 2D analysis of the cracked rock surface itself. The fragments collected at the base of the outcrop weathered from this surface.

We performed our standard 2D image analysis: after identifying the fracture network we counted the  degree of each cell 
(Figure \ref{fig:2D_Harmashatar} (a,b,e)) and each node  (Figure \ref{fig:2D_Harmashatar} (a,b,d)) to obtain the distributions $p(n), p(v)$ with averages $\bar v=4.19, \bar n=2.76$ (Figure \ref{fig:2D_Harmashatar} (d,e)). For the latter we separately counted the regular and irregular ones to obtain the regularity parameter $p$, and subsequently used formula (1) to determine $\bar v$ from the values of $\bar n$ and $p$. By comparing the two obtained values of $\bar v$, we tested the consistency of our image analysis (Figure \ref{fig:2D_Harmashatar} (f)).

As here we deal with a 2D mosaic, we only investigate planar geometric features without making any assumptions about the third dimension. As a consequence, the mass $m$ of fragments was determined as the number of pixels covered by the fragment area, defined as a region enclosed by intersecting fractures.
Figure \ref{fig:2D_Harmashatar} (c)  presents the probability distribution $p(m)$ of the mass of fragments  on a double logarithmic plot, where $m$ is normalized by the average mass of fragments. 
It can be observed that although $m$ spans a broad range, the distribution $p(m)$
is not power law. We expect that the fracture pattern was 
created by a sequential fracturing process, which usually leads to a power-law size distribution
of pieces. We suggest that the absence of a well-defined power law in Figure \ref{fig:2D_Harmashatar} (c) means that the fracture process is not fully developed, in the sense that we observe
only the first few generations of sequential breaking. Further sequential fragmentation, or an expanded domain of fractured rock (Fig. 2 of main text), would be expected to make the data converge to a power-law mass distribution.

The 2D analysis of the rock surface confirmed our hypothesis that the fragmentation pattern can be well approximated by an irregular primitive mosaic. The measured average value for the cell degree,
$\bar v=4.19$, is close to cubic and confirms that this mosaic is in the vicinity of the Platonic attractor.

\begin{figure}[ht]
\begin{center}
\includegraphics[width=0.7\columnwidth]{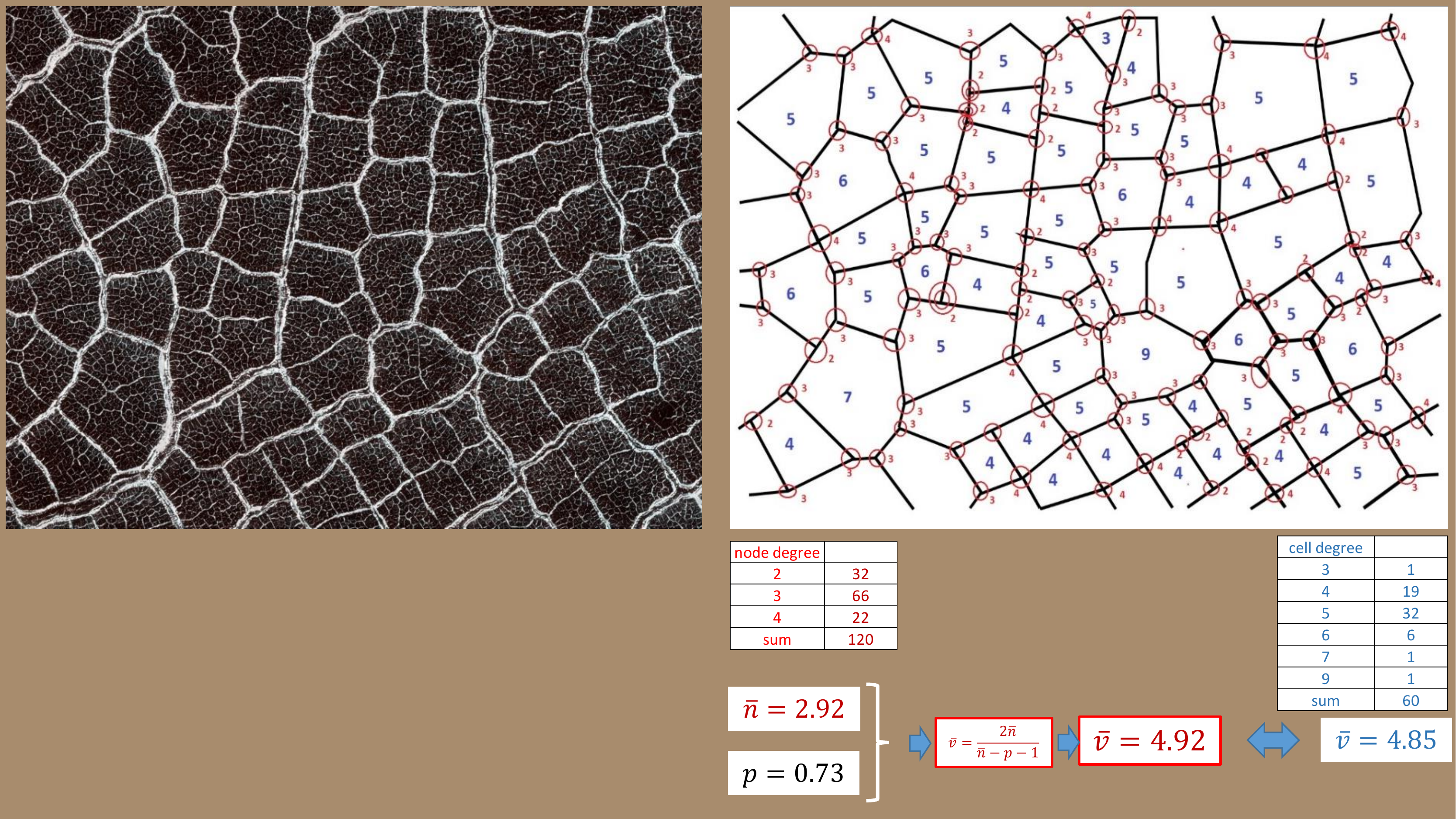}
\caption{\textbf{Analysis of Martian landscape.} Left: original image
\\
(\url{https://mars.nasa.gov/resources/5314/polygonal-patterned-ground/}, Image credit:  NASA/JPL-Caltech/University of Arizona). Upper right: 2D mosaic extracted. Blue numbers indicate cell degrees, red numbers indicate nodal degrees. Lower right: computation of averages. Average vertex degree has been computed both directly from the individual vertex degrees as $\bar v=4.85$, and from the nodal degrees via the general formula for planar mosaics as $\bar v = 4.92$. Good agreement between these numbers indicates internal consistency of our measurements, and that the data are well approximated by a convex mosaic. }
\label{fig:2D_nature}
\end{center}
\end{figure}

\begin{table} [htp] 
{
\begin{tabular}{| l |p{3.5cm} | p{12.5cm} |}
\hline
ID & Name & Source  \\
\hline
\hline
13 & Giant's Causeway, UK &
\begin{verbatim}
https://www.rob-tomlinson.com/places/giants-causeway
\end{verbatim} \\
\hline
 14 & Mud cracks & \begin{verbatim}
http://www.pitt.edu/~cejones/GeoImages/5SedimentaryRocks/
 SedStructures/Mudcracks.html\end{verbatim} \\
 \hline
 15 & Tectonic Mosaic & Image from \cite{muller2008age} \begin{verbatim}https://commons.wikimedia.org/wiki/File:
 Age_of_oceanic_lithosphere.jpg  \end{verbatim} \\
\hline
 16 & Martian landscape    & Photo: NASA/JPL-Caltech/University of Arizona \begin{verbatim} https://mars.nasa.gov/resources/5314/polygonal-patterned-ground/ \end{verbatim}		 \\
\hline
17 & Polygonal cracking on the North Slope of Alaska & Photo by M. Druckenmiller \begin{verbatim}https://astrobob.areavoices.com/2008/05/26/
 polly-want-a-polygon/ \end{verbatim}\\
\hline
18 & Mud cracks & Photo by H. Grobe. \begin{verbatim}https://en.wikipedia.org/wiki/File:
 Desiccation-cracks_hg.jpg
\end{verbatim}  \\
\hline
19 & Dolomite outcrop on H\'armashat\'arhegy, Hungary & Photo by the authors. \\
\hline
20 & Periglacial polygons on the North Slope of Alaska &
\begin{verbatim}http://www.groundtruthtrekking.org/photo/
 grid-like-polygonal-ground/
\end{verbatim} \\
\hline
21 & Granite outcrop, Helsinki, Finland & Photo by the authors \\
\hline
\hline
A & Septarian Nodule, 3D Voronoi &
\begin{verbatim}
https://www.fossilera.com/sp/128547/
 dragon-egg-geodes/septarian-with-black-calcite.jpg
\end{verbatim} \\
\hline
B & Giant's Causeway, UK, Columnar Voronoi &
\begin{verbatim}
https://www.rob-tomlinson.com/places/giants-causeway
\end{verbatim} \\
\hline
 C  & 3D cracked rock, 3D primitive & \begin{verbatim}	
http://maps.unomaha.edu/Maher/STEP07/supportinfo/cracks.html
 \end{verbatim} \\
\hline

\end{tabular}
}
\normalsize
\vspace{0.5cm}
\caption{\textbf{Sources of images of Fig.~3 and 5 of the main text}. ID: (numbers) reference number of the image in Fig.~3, (letters) reference letter of the image in Fig.~5 of the main text, Name: Description and location of the image, Source: URL locations of the images
}
\label{tab:2D_sources}
\end{table}

\subsection{The Tectonic Mosaic}\label{SI4}

We examine the tectonic plate configuration of the Earth \cite{bird2003updated} as a 2D convex mosaic (to which we will refer as the Tectonic Mosaic), treating the Earth's crust as a thin shell. According to \cite{bird2003updated}, there are 52 plates, 100 triple junction points and  150 plate boundaries (the latter are defined as the lines separating two plates). In the terminology of convex mosaics this translates into having $N_C=52$ cells, $N_V=100$ nodes of degree $n=3$ and 150 edges on a spherical 2D mosaic. 
 In 2 dimensions, the average number of edges associated with a cell equals the average number of vertices. Since each edge belongs to 2 cells, in the case of the Tectonic Mosaics the average number of vertices (the vertex degree) associated with one cell is $\bar v_{tectonic} = 300/52=5.769$. Since all nodes are degree 3, we have a regular mosaic. However, if we substitute $p=1, \bar n=3$ into formula (1), 
 instead of $5.769$ we obtain $\bar v=6$.  As we explain below, the difference $\bar v-\bar v_{tectonic}=0.231$ is the result of the Gaussian curvature of the Earth's surface.
 
\begin{figure}[ht]
\begin{center}
\includegraphics[width=1.0\columnwidth]{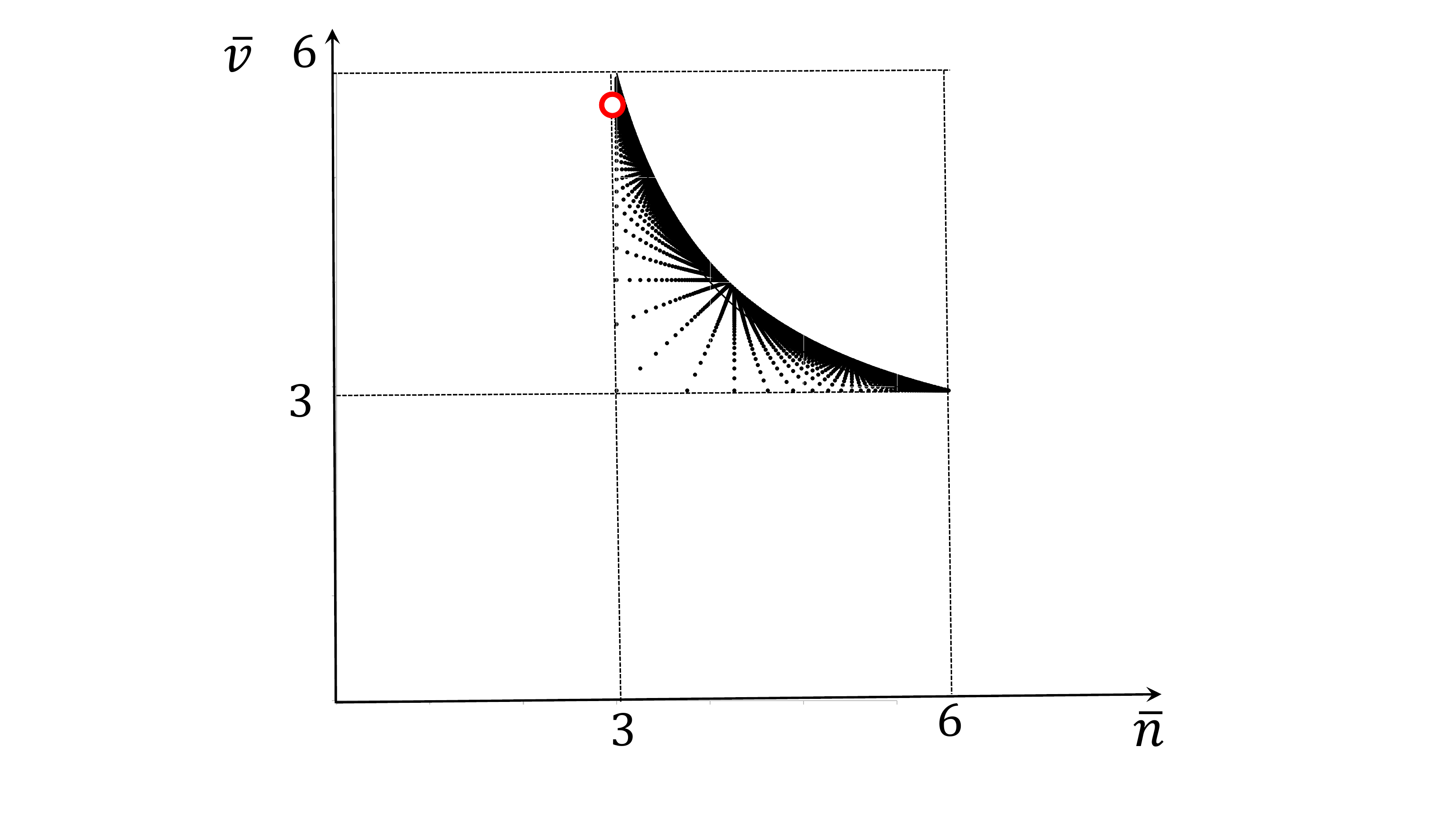}
\caption{\textbf{Regular 2D spherical mosaics on the symbolic $[\bar
n, \bar v]$ plane \cite{domokos2019honeycomb}.} Note that, unlike Euclidean mosaics, spherical mosaics do not form a continuum; however, with increasing number of vertices and faces, they accumulate on the $p=1$ line corresponding to Euclidean mosaics. This indicates that  for sufficiently high values of $N_v, N_f$, locally the cells are well approximated by Euclidean polygons. Observe the Tectonic Mosaic (marked with red circle) close to this line. The $\bar n=3$ line carries mosaics corresponding to simple polyhedra with an even number of vertices of order 3,  and the $\bar v=3$ line carries mosaics corresponding to simplicial polyhedra with an even number of triangular faces. The point $(\bar n, \bar v)=(3,3)$ corresponds to tetrahedral mosaics and their isomorphs. }\label{fig:SI_tectonic}
\end{center}
\end{figure}

Formula (1) of the main text has been derived for 2D mosaics with zero curvature. 
In two dimensions, spherical mosaics may be characterized by the \emph{angle excess} associated with their cells which is equal to the solid angle subtended by the cell or, alternatively, the spherical area of the cell. Note that for mosaics on the Euclidean plane the angle excess is
identically zero. Let $\mathcal{M}$ be a convex  mosaic on $\mathbb{S}^2$ (i.e.  a mosaic on a topological sphere, briefly, a spherical mosaic) with $N_v$ nodes and $N_c$ cells. Since $\mathcal{M}$ is a tiling of $\mathbb{S}^2$, the average area of a cell is $\Omega_C = \frac{4\pi}{N_c}$. Similarly, the average area of a cell in the dual mosaic is $\Omega_N = \frac{4\pi}{N_v}$.
For any spherical mosaic $\mathcal{M}$,  the quantity 
\begin{equation}\label{excess}
\bar{\mu}(\mathcal{M})= \frac{1}{\pi} \frac{\Omega_C \Omega_N}{\Omega_C + \Omega_N}
\end{equation}
has been called \cite{domokos2019honeycomb} the \emph{harmonic angle excess} of $\mathcal{M}$. Substituting the values $N_C=52$, $N_V=100$ into (\ref{excess}) yields  
\begin{equation}\label{tectonicexcess}
\bar \mu_{tectonic}=0.026
\end{equation}
for the Tectonic Mosaic. The harmonic angle excess is linked \cite{domokos2019honeycomb} to the harmonic degree $\bar h$ of the mosaic (defined in Section 2 of the main text) via
\begin{equation}
\bar h = 2 - \bar \mu.
\end{equation}
In \cite{domokos2019honeycomb} formula (1) is generalized to 2D spherical mosaics by using the concept of harmonic angle excess:
\begin{equation}\label{eq:spherical_nff}
\bar{v} = \frac{(2-\bar \mu)\bar{n}}{\bar{n}+\bar \mu -1-p}.
\end{equation}
If we substitute (\ref{tectonicexcess}) into (\ref{eq:spherical_nff})
along with the values $\bar n=3, p=1$ then we obtain the exact value $\bar v_{tectonic}=5.769$, illustrating that the Tectonic Mosaic is a spherical Voronoi tessellation. The small value of $\bar \mu$ is an indication of the relatively large number of tectonic plates. In the limit of infinitely many plates (if the relative size of plates remains bounded) we would approach the $\bar \mu =0$ case, where Euclidean geometry serves a valid local approximation for each plate. Figure \ref{fig:SI_tectonic} illustrates the Tectonic mosaic among all spherical mosaics and also illustrates how the latter converge to the Euclidean case.

\section{3D  mosaics}\label{SI5}

Here we provide some background on the geometric examples of 3D mosaics that we created to showcase the range of possible patterns. Unlike in $d=2$ dimensions, there is no general formula known to define all regular convex mosaics in the $[\bar n, \bar v]$ symbolic plane. The trivial prismatic extensions of 2D mosaics populate the $\bar n \in [6, 12]$ segment of the $\bar v=4\bar n /(\bar n -4)$ curve (corresponding to harmonic degree $\bar h = 4$). Some of these prismatic mosaics qualify as \emph{uniform honeycombs}, i.e convex mosaics which have convex uniform polyhedra as cells. The result that 28 such tessellations are possible is rather recent \cite{grunbaum1994uniform}.  Table \ref{tab:3D} provides a complete list of these mosaics and their duals. In addition, we also listed Poisson-Voronoi, Poisson-Delaunay and primitive (hyperplane) mosaics \cite{schneider2008stochastic} and we provided the values of $\bar n, \bar v, \bar f$ for each mosaic.  

\begin{figure}[ht]
\begin{center}
\includegraphics[width=0.95\columnwidth]{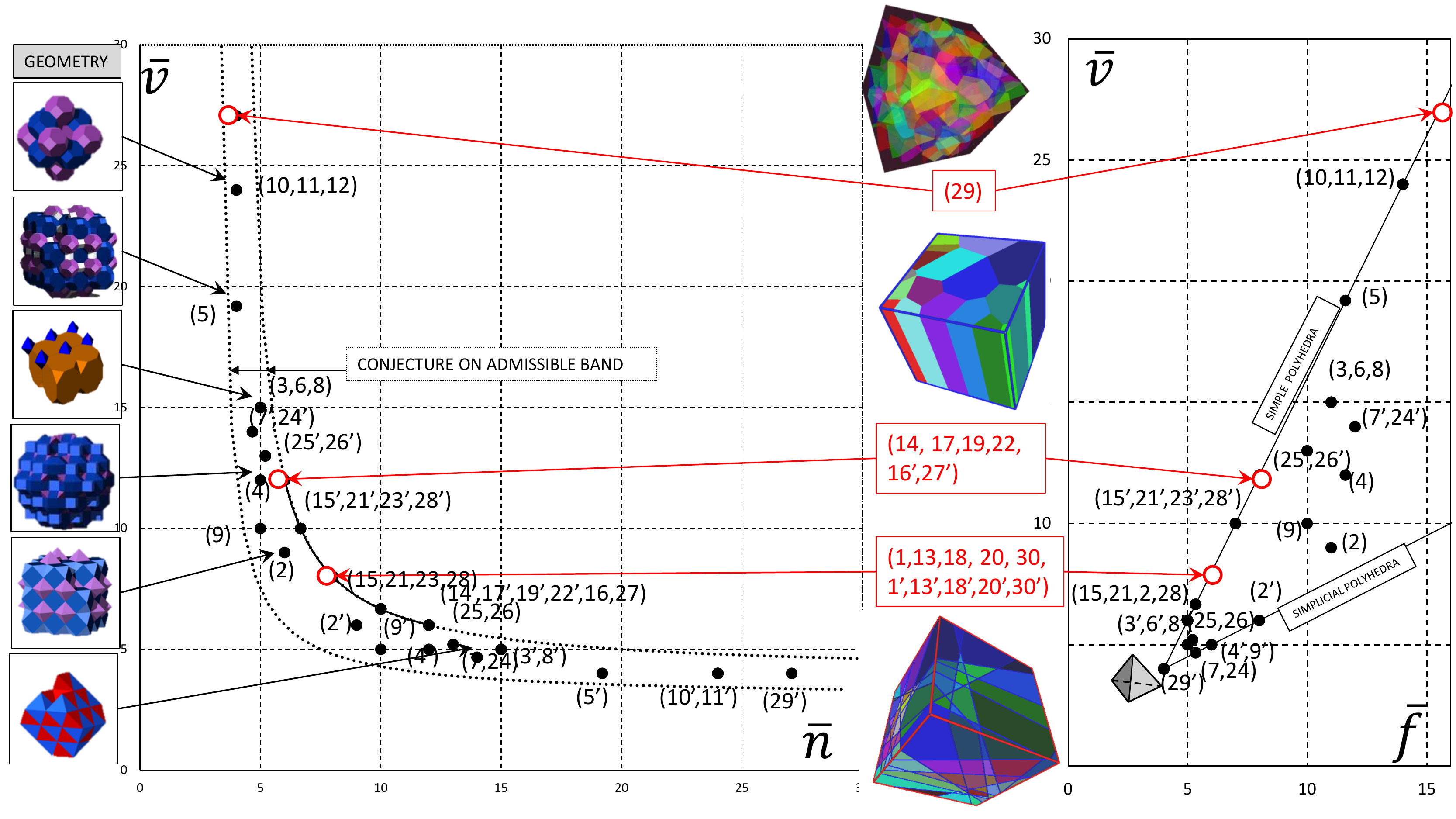}
\caption{\textbf{Illustration of the 3D mosaics}. (a) The $\bar v,~\bar n$ plane, and (b) the $\bar v,~\bar f$ plane. The mosaics in Table \ref{tab:3D} are marked with their respective row numbers. The three main observed fracture patterns are marked with red circles and additionally illustrated by insets. Solid line corresponds to prismatic mosaics, the dashed lines corresponds to 
conjectured bounds at $\bar h =3,4$, respectively, where $\bar h = (\bar n \bar v)/(\bar n + \bar v)$ is the harmonic degree of the mosaic \cite{domokos2019honeycomb}.}
\label{fig:SI_fig_3D_geometry}
\end{center}
\end{figure}

Figure \ref{fig:SI_fig_3D_geometry} illustrates these mosaics on the $[\bar n, \bar v]$ symbolic plane and the $[\bar f, \bar v]$ Euler plane. Each mosaic is identified with the serial number from Table \ref{tab:3D}. Red colors indicate mosaics with nodal and vertex degrees that have been also observed in nature. Some selected mosaics are illustrated with simple plots.

\begin{table} [htp] 
{\footnotesize	
\begin{tabular}{| c |r||c|c|c|r|}
\hline
ID.  & Name of mosaic & $\bar n$ & $\bar v$ & $\bar f$ & $\bar h$  \\
\hline
\hline
 1 & cubic    & 8 & 8 & 6 & 4.00		 \\
\hline
2 & rectified cubic	& 6 & 	9	& 11	& 3.60	 \\
\hline
3 & truncated cubic &	5	& 15	& 11	& 3.75	 \\
\hline
4 & cantellated cubic & 	5	& 12 & 	11.6 &	3.53	 \\
\hline
5 & cantitruncated cubic	& 4	& 19$\frac{1}{5}$&	11$\frac{3}{5}$ &	3.31	 \\
\hline
6 & runcitruncated cubic	& 5	 & 15	& 11	& 	3.75	\\
\hline
7 & alternated cubic	 &  14	 & 4$\frac{2}{3}$ &	5$\frac{1}{3}$	& 3.50		 \\
\hline
8 & cantic cubic	 & 5	& 15 & 	11	&  3.75		 \\
\hline
9 & runcic cubic	& 5	& 10	& 10	& 3.33	 \\
\hline
10	 & runcicantic cubic &	4	& 24	& 14	& 3.43	 \\
\hline
11	 & bitruncated cubic	 & 4	& 24	& 14	& 3.43		 \\
\hline
12	& omnitruncated cubic & 	4	& 24	 & 14	& 3.43	\\
\hline
13	& quarter cubic & 	8	& 8	& 6	& 4.00		 \\
\hline
14	& truncated/bitruncated square prismatic &	6	& 12	& 8	& 4.00	 \\
\hline
15	& snub square prismatic & 	10	& 6$\frac{2}{3}$ &	5$\frac{1}{3}$ &	4.00	 \\
\hline
16	& triangular prismatic	& 12	& 6	& 5	&  4.00	\\
\hline
17	& hexagonal prismatic	& 6	& 12	& 8	& 4.00		 \\
\hline
18	& trihexagonal prismatic &	8	& 8	& 6 & 	4.00	 \\
\hline
19	& truncated hexagonal prismatic	& 6	& 12 &	8	&  4.00	 \\
\hline
20	& rhombi-hexagonal prismatic	& 8	& 8	& 6	&  4.00		 \\
\hline
21		& snub-hexagonal prismatic	& 	10	& 	6$\frac{2}{3}$	& 	5$\frac{1}{3}$		& 4.00	 \\
\hline
22		& truncated trihexagonal prismatic		& 6		& 12		& 8		&  4.00	 \\
\hline
23		& elongated triangular prismatic		& 10		& 6$\frac{2}{3}$	& 5$\frac{1}{3}$	& 4.00		 \\
\hline
24		& gyrated alternated cubic		& 14		& 4$\frac{2}{3}$	& 5$\frac{1}{3}$		&  3.50	 \\
\hline
25		& gyroelongated alternated cubic		& 13		& 5$\frac{1}{5}$	& 5$\frac{1}{5}$		& 3.71	 \\
\hline
26		& elongated alternated cubic		& 13		& 5$\frac{1}{5}$		& 5$\frac{1}{5}$ &  3.71	 \\
\hline
27		& gyrated triangular prismatic		& 12		& 6		& 5		&  4.00	 \\
\hline
28		& gyroelongated triangular prismatic		& 10		& 6$\frac{2}{3}$		& 5$\frac{1}{3}$		& 4.00		 \\
\hline
29 & 	Poisson-Voronoi	& 4	& 27.07 &	15.51 &	 3.49	 \\
\hline
30	& Primitive &	8	& 8	& 6	&  4.00		 \\

\hline
1'	& dual of cubic	& 8	& 8	& 6	& 4.00		 \\
\hline
2'	& dual of rectified cubic &	9 &	6	& 8	&  3.60		 \\
\hline
3'	& dual of truncated cubic	& 15	& 5	& 5	&  3.75		 \\
\hline
4'	& dual of cantellated cubic	& 12	& 5	& 6	& 3.53		 \\
\hline
5'	& dual of cantitruncated cubic	& 19$\frac{1}{5}$	& 4 & 	4	&  3.31		 \\
\hline
6'	& dual of runcitruncated cubic	& 15	& 5 &	5	&  3.75	 \\
\hline
7'	& dual of alternated cubic	& 4$\frac{2}{3}$	& 14 &	12	&3.50		 \\
\hline
8'	& dual of cantic cubic &	15	& 5	& 5 &	  3.75	 \\
\hline
9' & 	dual of runcic cubic	& 10	& 5	& 6	& 3.33	 \\
\hline
10'	& dual of runcicantic cubic &	24	& 4	& 4	& 3.43	 \\
\hline
11'	& dual of bitruncated cubic	& 24	& 4	& 4	& 3.43		 \\
\hline
12'	& dual of omnitruncated cubic	& 24	& 4	& 4	& 3.43		 \\
\hline
13'	& dual of quarter cubic	& 8	& 8	& 6	& 4.00		 \\
\hline
14'	& dual of truncated/bitruncated square prismatic & 	12	& 6 & 5 &   4.00		 \\
\hline
15'	& dual of snub square prismatic	& 6$\frac{2}{3}$ & 	10 &	7	& 4.00	 \\
\hline
16'	& dual of triangular prismatic &	6 &	12	& 8	& 4.00		 \\
\hline
17'	& dual of hexagonal prismatic &	12	& 6	& 5 & 4.00	 \\
\hline
18'	& dual of trihexagonal prismatic &	8	& 8	& 6	& 4.00		 \\
\hline
19'	& dual of truncated hexagonal prismatic & 	12	& 6	& 5	& 4.00		 \\
\hline
20'	& dual of rhombi-hexagonal prismatic &	8	& 8	& 6	& 4.00		 \\
\hline
21'	& dual of snub-hexagonal prismatic & 	6$\frac{2}{3}$ &	10	& 7	& 4.00		 \\
\hline
22'	& dual of truncated trihexagonal prismatic &	12	& 6 & 5 &	 4.00		 \\
\hline
23'	& dual of elongated triangular prismatic &	6$\frac{2}{3}$	& 10	& 7	& 4.00		 \\
\hline
24'	& dual of gyrated alternated cubic &	4$\frac{2}{3}$ &	14 &	12 & 3.50		 \\
\hline
25'	& dual of gyroelongated alternated cubic &	5$\frac{1}{5}$ &	13 &	10	&  3.71	 \\
\hline
26'	& dual of elongated alternated cubic &	5$\frac{1}{5}$ &	13	& 10	& 3.71	 \\
\hline
27'	& dual of gyrated triangular prismatic & 	6	& 12 &	8 &	 4.00		 \\
\hline
28'	& dual of gyroelongated triangular prismatic	& 6$\frac{2}{3}$ &	10 &	7	& 4.00	 \\
\hline
29'	& Dual of Poisson-Voronoi: Poisson-Delaunay &	27.07 &	4	& 4	&3.49		 \\
\hline
30'	& Dual of Primitive: Primitive &	8 &	8	& 6	&  4.00		 \\
\hline

\end{tabular}
}
\normalsize
\vspace{0.5cm}
\caption{\textbf{List of uniform convex honeycombs.} Nodal degree $\bar n$, vertex degree $\bar v$ and harmonic degree $\bar h$ of uniform convex honeycombs, their duals, Poisson-Voronoi, Poisson-Delaunay and Primitive (Hyperplane) random mosaics.}
\label{tab:3D}
\end{table}

\section{Discrete element method simulations}\label{SI6}
This section details the methods employed to produce discrete element method (DEM) fragmentation results that are presented in the main text. This includes model choices, and how model results were analyzed. This information shows how fracture mosaic patterns are related to the formative stress field. We also cast our results in terms of the classic Flinn diagram that is typically used in geology and rock mechanics literature.

\subsection{Computational model} \label{ss:SI_protocols}

In order to compute an approximation of the pattern generator (i.e. to establish the link between stress fields and fragmentation patterns) we have used a modified version of the LAMMPS \cite{plimpton1995fast} discrete element method (DEM) simulation package which included breakable glued contacts between particles\cite{brendel2011contact}. The initial sample was prepared by the following procedure (using arbitrary time, mass and length units):
\begin{itemize}
    \item $N$ spherical particles with diameter uniformly distributed on $[0.9,1.1]$ were placed in a periodic cube (or square in two dimensions) using random sequential deposition.
    \item The (periodic) cube was contracted using Hertzian particle contact model until reaching  the average of $6$ contacts per particle ($4$ contacts per particle in two dimensions).
    \item After the initial configuration was established by the above protocol, existing contacts were glued together while the stresses are removed from the system. The simulation started by applying displacement (strain) control,
    using pre-defined strain levels in the 3 orthogonal directions.
    \item Displacement of box boundaries was always parallel.
    \item During the time evolution of the particle system, the boundary velocities were kept constant.
    
\end{itemize}
 The  parameters of the simulations were the following: 
 \begin{itemize}
     \item  Hertzian elastic constant $k_n{=}10^5$,
     \item  density of the particles $\rho =1$,
     \item  microscopic friction coefficient between particles $\mu{=}0.5$,
     (which resulted in a macroscopic effective friction coefficient of $\mu_{eff}\simeq 0.6-0.9$.)
     \item  timestep  $\Delta t{=}10^{-4}$,
     \item number of particles $N{=}6\cdot10^5$ (3D, large simulation);  $N{=}2\cdot10^5$ (3D, small simulation),
     \item strain levels $\epsilon = \pm (0,~0.005,~0.01,~0.015,~0.02)$
     \item Young's modulus $E=10^3$ Pa (soft material \cite{brendel2011contact}),
     \item Poisson's ratio $\mu =0.3$.
 \end{itemize}

In case of tensile strain one single system-wide crack is sufficient to release the stresses in the system, so we set a relatively large ($\eta =1000$) viscous friction between the particles and the boundaries; this gave rise to reasonable sized fragments.
 For a better visualization, after the simulation is completed the sample was expanded to increase the opening of cracks.

\subsection{Results}\label{ss:results}

\subsubsection{Large simulations: main patterns}
Three test samples with $N=5\cdot 10^5$ particles were investigated to reproduce the main observable 3D patterns shown in the middle column of Fig.~\ref{fig:SI_fig_3D_geometry}. The resulting mosaics are shown in Fig.~5 of the main text. We created the 3D Voronoi mosaic by uniform tension (strains $0.01,0.01,0.01$). The columnar mosaic
was created with uniform tension in two perpendicular directions, while in the third direction we applied a compression to compensate the contraction according to the Poisson's ratio
(strains $0.01,0.01,-0.003$, the latter has been used only in this computation). The 3D primitive mosaic was created by applying different negative strains ($-0.02,-0.01,-0.005$) in each direction.

\subsubsection{Small simulations: the 3D pattern generator}

In addition to the three large simulations, we also performed $90$ smaller simulations with $N{=}2\cdot 10^5$ of very soft particles ($E{=}1$ kPa). The goal
of these simulations was to scan the $[\mu_1,\mu_2]$, $i=\pm 1$ planes
to obtain a picture of the 3D pattern generator --- albeit a crude one.
We used the simulation parameters and protocols described in subsection \ref{ss:SI_protocols}.  We adjusted strain values to obtain an approximately equidistant mesh with $\Delta \mu =0.25 $ in both directions, on both the $i=1$ and the $i=-1$ planes of the $[\mu_1,\mu_2]$ map.

After completing the tests we
constructed 3 orthogonal sections of the final configuration of the sample and identified visually the resulting junction types. Planar sections of a 3D Voronoi mosaic display ``Y'' junctions on all planar cuts. Columnar mosaics have one principal plane with only ``Y'' junctions, while the other directions may only contain both  ``T'' and ``X'' junctions. Primitive mosaics may not contain ``Y'' junctions at all. For some stresses, several independent simulations were executed. If the resulting crack network was inconclusive or if individual runs showed different features, we marked the simulations for both mosaic types.

\begin{figure}[ht]
\begin{center}
\includegraphics[width=0.95\columnwidth]{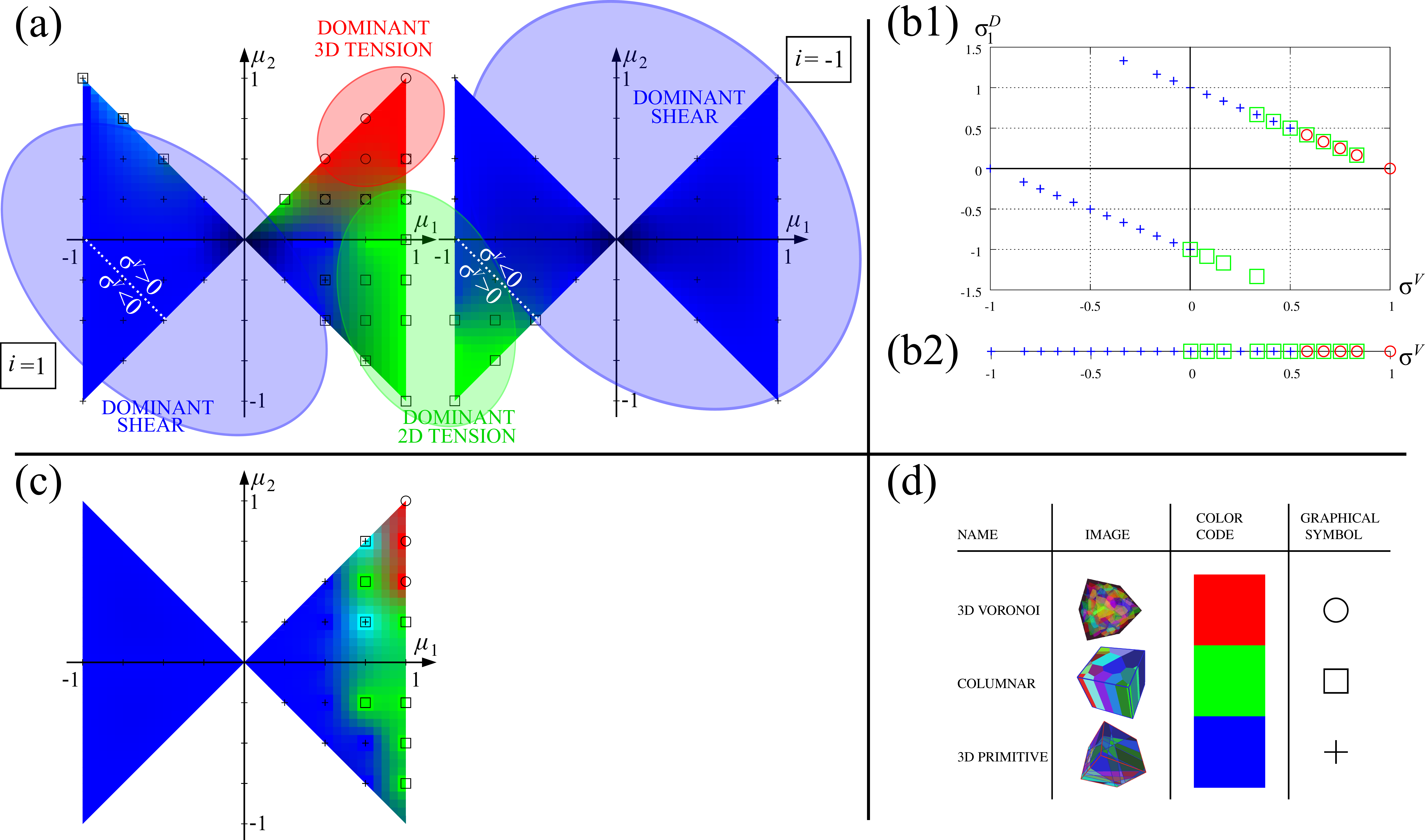}
\caption{\textbf{The 3D pattern generator} (a) The two leaves of the $[\mu_1,\mu_2]$ map for $i{=}1$ on the left and $i{=}-1$ on the right for simulations with $E{=}1$kPa. For a given set of $[\mu_1,\mu_2,i]$ the resulting fracture patterns are marked with symbols; coloring was made using Laplacian approximation. Colors and symbols defined in the legend in part (d) of the figure. (b1) Observed patterns in the plane of the volumetric-normalized deviatoric of the largest stress [$\sigma^V, \sigma_1^D$]. (b2) is the one dimensional version of (b1) using only the volumetric stress. (c) The $i{=}1$ leaf of the pattern generator map for simulations with $E{=}10kPa$. Similarly to panel (a) coloring was done with a Laplacian approximation. (d) Graphic and color codes for the three observed patterns.
}
\label{fig:DEM_butterfly}
\end{center}
\end{figure}

\subsection{Interpretation of results}
In Fig.~\ref{fig:DEM_butterfly} (a) we have plotted the different mosaic types with different symbols (see Fig.~\ref{fig:DEM_butterfly} (d) for the legend) for different values of $(\mu_1,\mu_2)$. The underlying map was colored according to the patterns observed, using Laplacian interpolation. 
As we can observe, 3D Voronoi mosaics (red) may be generated only by almost uniform tension. Columnar mosaics (green) are generated by two directions with tension and with low deviatoric stresses, while the rest of the available parameter space is covered by  primitive mosaics (blue).

Figure \ref{fig:DEM_butterfly}(a) shows that under compressive stresses, regular primitive mosaics are the \emph{only} observable pattern. In the presence of tension (cf. also \cite{clair2015geophysical}), columnar and 3D Voronoi mosaics may also emerge; however, we emphasize that even in this case only a suitably chosen deviatoric stress component will result in either columnar or 3D Voronoi  mosaics. 
This demonstrates that in the compression-dominated geologically relevant processes, only fragmentation patterns are obtained which may be well approximated by primitive mosaics.

To make these observations more definite, in Fig.~\ref{fig:DEM_butterfly} (b) we have plotted the first deviatoric stress $\sigma^D_1=\sigma_i-\sigma^V$  versus
the volumetric stress $\sigma^V=(\sigma_1+\sigma_2+\sigma_3)/3$.
Individual, marked points on the diagram correspond to the 90 small simulations
(discussed in Subsection \ref{ss:results}), using the color and graphical codes of Fig.~\ref{fig:DEM_butterfly} (d). This
representation allows for clear criteria separating the three main patterns:

\begin{equation}
    \begin{cases}
    \sigma^D_1>0&
    \begin{cases}
    \sigma^V<0 & \mathrm{primitive}\cr
    \sigma^V>0 & \mathrm{columnar}
    \end{cases}\cr
    \sigma^D_1<0&
    \begin{cases}
    \sigma^V<1/4 & \mathrm{primitive}\cr
    1/4<\sigma^V<1/2 & \mathrm{primitive/columnar}\cr
    1/2<\sigma^V & \mathrm{columnar/Voronoi}\cr
    1\simeq\sigma^V & \mathrm{Voronoi}
    \end{cases}
    \end{cases}
\end{equation}

While regular primitive mosaics dominate all plots of Figure \ref{fig:DEM_butterfly} (a), the domains occupied by  columnar and 3D Voronoi mosaics are still significant. However, the extent of these domains  very much depends on the softness/hardness of the material considered: softer material
will result in larger stress-domains occupied by columnar and 3D Voronoi mosaics. Our simulations were performed with a material which is orders of magnitude softer compared to most geomaterials, so we expect the relative significance of columnar and 3D Voronoi mosaics to be significantly smaller in natural processes compared to what Figure
\ref{fig:DEM_butterfly} suggests. To support this observation, additional simulations were carried out with materials having a ten-times higher stiffness ($E{=}10$kPa). As we can observe on panel (c) of Figure \ref{fig:DEM_butterfly}, harder material (much closer to realistic geomaterials) allows only a considerably smaller area for possible columnar and Voronoi mosaics.

\subsection{The Flinn diagram}\label{sec:flinn}

In the geological literature, when general deformation patterns are studied the deformation state is often presented on the so-called \emph{Flinn diagram}
\cite{flinn1962diagram} which illustrates the deformation of an infinitesimal sphere into an ellipsoid with principal semi axes $\lambda_1{\geq}\lambda_2{\geq}\lambda_3$ by plotting $\log(\lambda_1/\lambda_2)$ versus $\log(\lambda_2/\lambda_3)$. 
For better comparison, we will illustrate the result of our simulation also on this diagram.

\begin{figure}[ht]
\begin{center}
\includegraphics[width=0.7\columnwidth]{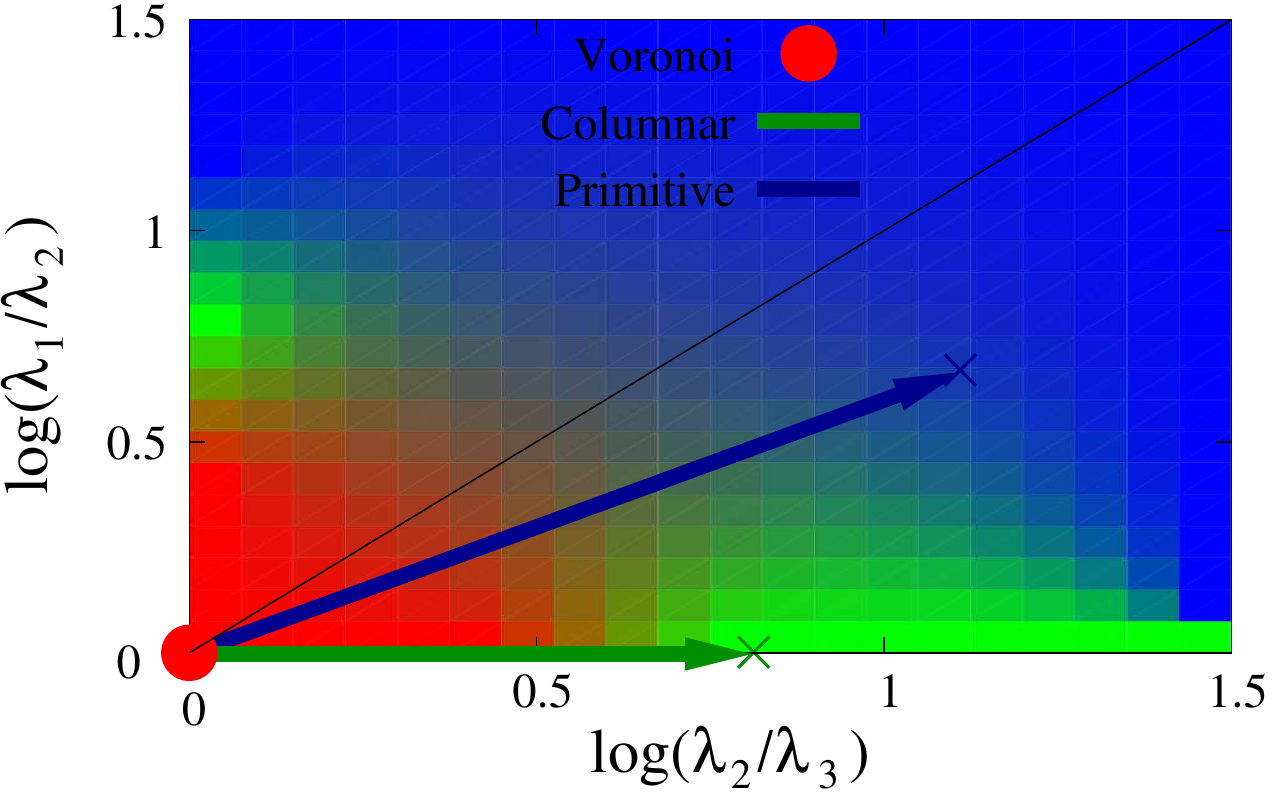}
\caption{\textbf{Flinn diagram.} Flinn diagram displaying the simulation path of the 3D DEM simulations presented in the main text in Figure~5. Here, the primitive mosaics associated with positive strains are shown, as opposed to the one in Figure 5. with negative strains; the latter could not be depicted in this figure. Positions illustrated by the insets in Figure 5 are marked with an "x".}
\label{fig:flinn}
\end{center}
\end{figure}

As the Flinn diagram displays deformation, a simulation may be fully represented by a curve rather than a single point, so it appears less convenient than our stress-based representation in Figure 5 of the main article and Figure \ref{fig:DEM_butterfly}. In the Flinn diagram the representation of relative strains is non unique --- in close analogy with the $i=\pm 1$ ambiguity of our stress-based representation. However, since geologically relevant scenarios are mostly dominated by compression, this non-uniqueness rarely leads to misunderstandings. In Figure \ref{fig:flinn} we plotted  the geologically less relevant  \emph{tension} leaf of the Flinn diagram
(analogous to the $i=1$ leaf of our stress representation) and depicted our computer simulations in this diagram.
Our large scale simulations (represented by the snapshots in Fig.~5 of the main paper) are shown with the full corresponding trajectories; for easier illustration, as opposed to the example
shown in Fig 5. of the main text, here we show a regular primitive mosaic corresponding to positive strains. The three main mosaics are shown as follows:

\begin{itemize}
    \item The 3D Voronoi mosaic, corresponding to hydrostatic tension, is located at the origin of the Flinn diagram.
    
    \item Columnar mosaics are created if two positive principal stresses (and thus the corresponding deformations) are identical. In the Flinn diagram only the axes fulfill this requirement and indeed the trajectory of our columnar simulation lies on the horizontal axis. Points on the vertical axis may, in principle, correspond also to columnar mosaics; however, as we can see from Figure~\ref{fig:DEM_butterfly}, large tensile stress may lead to primitive mosaics. 
    
    \item Regular primitive mosaics (marked with blue) can be found at generic locations of the Flinn diagram.  A point in the graph can be represented by the tangent of the angle $K$ and the distance from the origin $D$. The shear pattern is in general determined by $K$ and the intensity or density of the pattern by $D$. The regular primitive pattern marked by "X" corresponds to
    positive strains (as opposed to the one illustrated in in Figure 5 of the main text).
\end{itemize}

The smaller simulations are shown by the same color code we used in Figure \ref{fig:DEM_butterfly}.
We can observe that the ratio of domains occupied by 3D Voronoi, columnar and regular primitive mosaics appears very similar to the $i=1$ leaf of the stress-based representation in Figure \ref{fig:DEM_butterfly}(a).
We also note that the other leaf of the Flinn diagram (corresponding to purely compressive principal strains, not shown in Figure \ref{fig:flinn}) would appear similar to the $i=-1$ leaf in Figure \ref{fig:DEM_butterfly}(a), containing almost exclusively regular primitive mosaics.

Flinn diagrams appear to offer the same general insight regarding the relative weight of fragmentation patterns as our stress-based representation: they illustrate that among primary fragmentation patterns, regular primitive mosaics are the most common ones. This shows that, even without the effects of secondary fracture, already primary patterns are dominated by Platonic averages. 

Flinn diagrams have the advantage that they may help to link these ideas with traditional geological intuition. Our stress-based system offers the theoretical advantage that the mapping from the space of stress ratios into the space of mosaics is unique, and it defines the universal pattern generator. Our simulations are far from providing full information on this fundamental function; nevertheless, the computational results admit a good qualitative insight.

\section{Validation of the theory of primitive mosaics: geometric simulations fit to a large experimental dataset}
\label{sec:validate}
In this section we describe field measurements of natural fragments which are used to provide a test of the `cuboid average' predicted for 3D rock fragments under generic stress conditions. We then develop the computational models for regular and irregular primitive mosaics to rationalize the dominance of the `Platonic attractor' (cuboid averages) in 3D and to explain the full distribution of shapes of natural rock fragments. First is the \textit{cut} model that serves as the simplest representation of a regular primitive mosaic (primary fracture), and second is the \textit{break} model what approximates
an irregular primitive mosaic (secondary fracture). The statistical distributions obtained from these models are fitted to field data using a minimal number of parameters, and are found to explain the distribution of fragment shapes and shape descriptors very well.

\subsection{Field measurements}

We have collected experimental samples of different origins to validate our predictions. Data sources are summarized in Table \ref{tab:expsource}.  We used mainly the shape descriptors in Table \ref{tab:p2}, with the following additions/modifications:
(a) we also recorded the mass $m$; and (b) axis ratios were not computed based on the principal lengths $a\geq b \geq c$ of the smallest bounding box, but rather we measured the principal length
$\bar a \geq \bar b \geq \bar c$ following standard geological field protocols where $\bar a$ is the largest diameter, $\bar b$ is the largest diameter orthogonal to $\bar a$ and $\bar c$ is the diameter orthogonal both to $\bar a$ and $\bar b$. To distinguish between the two protocols, we denote axis ratios from the second (geological standard) protocol by $y_1,y_2$ (as opposed to $Y_1,Y_2$ in Table \ref{tab:p2}). We remark that 
the geological standard protocol produces identical results with the one in Table \ref{tab:p2} if the measured object is ellipsoidal. Differences become more significant for polyhedral shapes. If our goal had been to prove the existence of cuboid averages then the protocol in Table \ref{tab:p2} would have been the preferred one. However, this was \emph{not} our goal: as we pointed out in subsection \ref{ss:cuboid_descriptors}, we do not expect to recover cuboid averages among axis ratios. Rather, we only used axis ratios to test the correspondence and match between model and field data, and the standard geological protocol is suitable for this purpose.
We used two sets of experimental samples. For the \emph{small sample} (sample 1) with 556 fragments, we have 7 shape descriptors available ($m, f,v,y_1, y_2, S,U$). For the \emph{large sample} (samples 2-6) with altogether 3728 fragments, we have 5 shape descriptors ($m, y_1, y_2,S,U$) available. Data are summarized in Table \ref{tab:expsource}.

\begin{table*}[]
    \centering
    \begin{tabular}{|c|c|c|c|c|c|c|}
    \hline
          & Type & Location  & Material &  Num. &   [mm] & recorded   \\
         \hline
         \hline
         1 & Weathering &  Budapest & Dolomite & 556 & 25 - 120 & $m,S,U,y_1,y_2, f,v$ \\
         \hline
         2 & Hammering & Lab & Gypsum & 314 & 15 - 130 & $m,S,U,y_1,y_2$ \\
         \hline
           3 & Explosion & Keszthely & Dolomite & 2000 & 16 - 131 & $m,S,U,y_1,y_2$ \\
         \hline
           4 & Explosion & K\'ad\'arta & Dolomite & 1000 & 17 - 136 & $m,S,U,y_1,y_2$ \\
         \hline
           5 & Hammering & Budapest & Dolomite & 200 & 17 - 69 & $m,S,U,y_1,y_2$ \\
                    \hline
           6 & Hammering & Zakynthos & Limestone & 214 & 14 - 108 & $m,S,U,y_1,y_2$ \\
         \hline
    \end{tabular}
    \caption{\textbf{Experimental datasets used in the paper.} Columns: 1. sample number; 2. the type of fragmentation process; 3. locations of the fragments where Budapest, Keszthely and Kádárta are located in Hungary, and Zakynthos is in Greece; 4. material of the fragments; 5. number of fragments; 6. size of the fragments; and 7. recorded data.}
    \normalsize
    \label{tab:expsource}
\end{table*}

\subsection{Geometric model for primitive mosaics}

\subsubsection{General description of models}

In this section we describe in detail the geometric models
for primitive mosaics referred to in the main text as the
\emph{cut} model and the \emph{break} model, aiming to describe regular and irregular primitive mosaics, respectively. (i) The \textit{cut} model aims to reconstruct the primary fragmentation pattern of a regular primitive mosaic: We start from a unit cube which is
cut by $N$ random planes. Planes are defined by a uniformly selected random point in the unit cube and by a uniformly selected random normal
direction. These subdivide the unit cube into convex polyhedra, which we regard as the geometric models of
fragments. We disregard the polytopes whose boundary overlaps with the boundary of the original unit cube, keeping only polytopes in the interior. (ii) The \textit{break} model is aimed to reconstruct the fragmentation pattern due to a sequential, binary breakup process. We implement the sequential breakup in the spirit of the model described in \cite{domokos_universality_2015}: the polyhedra resulting from the Cut model are subjected to binary breakup, with probability proportional to $1-y_1y_2$. If a polyhedron is broken into two parts, these secondary polyhedra are also subject in the next step to the same break dynamics.

\subsubsection{Parameter fitting: compensating for sampling bias}

In order to compare numerical results with the experimental data we have to take into account several sampling biases:
\begin{itemize}
\item There is always a lower cutoff in size for the experimental samples and we shall implement the same for our geometrical models. Thus we select only fragments the mass of which is larger than a pre-set value $m_0$.
\item The determination of marginally stable and unstable points of fragments contains some ambiguity. To compensate for any resulting inaccuracies,
we randomly varied the exact center of mass of the polyhedra using a normal distribution with variances proportional to the principal lengths: $(\sigma_0 a,\sigma_0 b,\sigma_0 c)$.
We kept only those equilibrium points which were identified in 95\% of these
randomly selected cases. 
\item The third effect considers the number of vertices and faces of the fragments. We assume that faces smaller than $A_0P$ will not be found by experimenters, where $P$ denotes the smallest projected area of the polyhedron. Making small faces disappear was implemented by letting the incoming edges be joined at a virtual vertex, admitting the existence of  non-simple polyhedra in our model.
\end{itemize}  
By letting the above three parameters ($m_0, \sigma_0, A_0$) vary, we fitted the density functions for $m, f,v,y_1,y_2, S,U$ of our seven shape descriptors by minimizing the largest deviation on all the listed seven density functions. We applied the fit to both the cut and break models for comparison. 

The general effects of our fitting parameters can be summarized as follows: The application of $m_0$ excludes small fragments,
which appear to be dominantly tetrahedra. This is intuitively clear: small polyhedra are being produced by chopping off vertices of larger polyhedra. Since all vertices are of degree 3, the small chopped-off fragment will be a tetrahedron in most cases. Objects with few faces and vertices can have few equilibria. In case of the large dataset the numbers $S,U$ have been systematically underestimated in the hand experiments. To compensate for this error the  value of $m_0$ had to be kept smaller for the large dataset.

The application of $A_0$ results in the creation of new vertices with degree higher than 3, so it produces polyhedra which are not simple (although theory predicts only simple polyhedra for primitive mosaics).

The value of $\sigma_0$ also decreases the number of equilibrium points. Although we do not have theoretical proof of this phenomenon, numerical simulations indicate that random polyhedra, on average, have the highest value for $S,U$ if the material is homogeneous.

\subsection{Interpretation of results}

We have thus four different sources of information: 
\begin{enumerate}
    
    \item mathematical theory of primitive mosaics;
    \item geometric model dataset (cut and break models) fitted
    to experiments; 
    \item small experimental dataset; and
    \item large experimental dataset;
\end{enumerate}
and four types of shape descriptors:
\begin{itemize}
    
    \item (a) numbers $f,v$ of faces and vertices;
    \item (b) axis ratios $y_1,y_2$;
    \item (c) numbers $S,U$ of stable and unstable static equilibria; and
    \item (d) mass.
\end{itemize}
Table \ref{tab:datasource} summarizes the shape descriptors available for the listed sources. Regarding (c)  we note that the protocols determining the values of $S$ and $U$ were slightly different between Samples 1 and 2; for Sample 1 corresponding to the small dataset, measurements were taken with much more care than the samples in the large dataset. This difference explains why we obtained slightly different optimal fitting parameters for the small and the large datasets.

\begin{table*}[ht]
    \centering
    \begin{tabular}{|r c||c|c|c|c|}
    \hline
      & & (a)  & (b)  & (c) &  (d)     \\
         \hline
         
        &  Source & $f,v$ & $y_1,y_2$  & $S,U$ &  $m$     \\
         \hline
           
         \hline
         1. & Theory & YES &  NO & NO & NO  \\
         \hline
         2. & Geometrical model & YES & YES & YES & YES \\
         \hline
         3. &  Small dataset  &  YES & YES & YES & YES \\
         \hline
         4. & Large dataset & NO & YES & YES & NO \\
         \hline
    \end{tabular}
    \caption{\textbf{Sources of data.}}
    \normalsize
    \label{tab:datasource}
\end{table*}

\begin{figure}[ht]
\begin{center}
\includegraphics[width=0.9\textwidth]{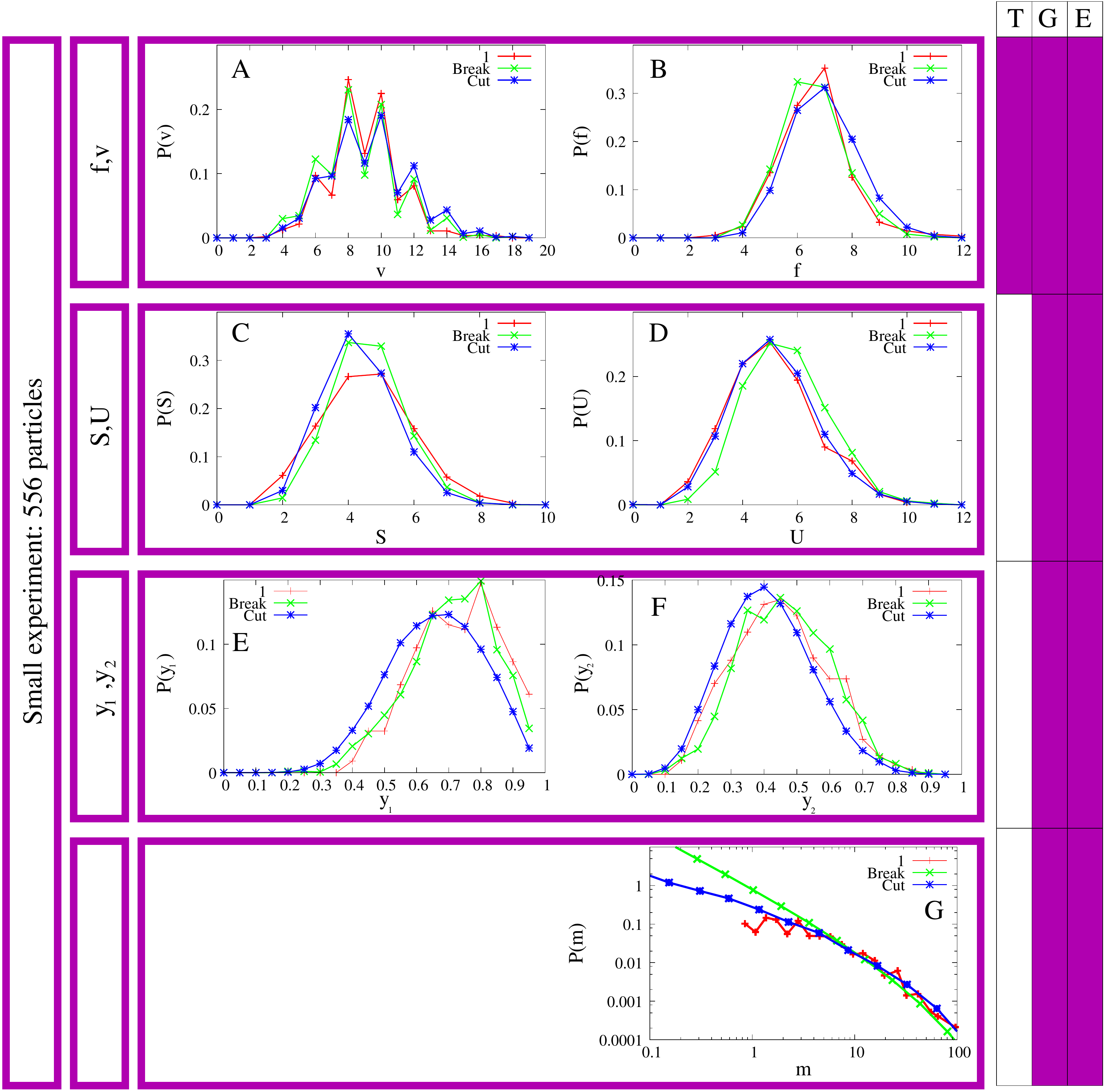}
\caption{\textbf{Fit of the small experimental dataset (Table \ref{tab:expsource}, 1) with the geometric model.} First row (A and B): vertices ($v$) and faces ($f$), Second row (C and D): stable equilibria ($S$) and unstable equilibria ($U$), Third row (E and F): axis ratios ($y_1,y_2$) , Fourth row (G): mass ($m$). Fitting parameters  $[m_0,\sigma_0,A_0]=[5\cdot10^{-4},0.05,0.1]$ for the cut and $[m_0,\sigma_0,A_0]=[10^{-4},0.05,0.1]$ for the break model; the associated $R_\mathrm{max}^2 \geq 0.99$.}
\label{fig:big1}
\end{center}
\end{figure}

\begin{figure}[ht]
\begin{center}
\includegraphics[width=0.9\textwidth]{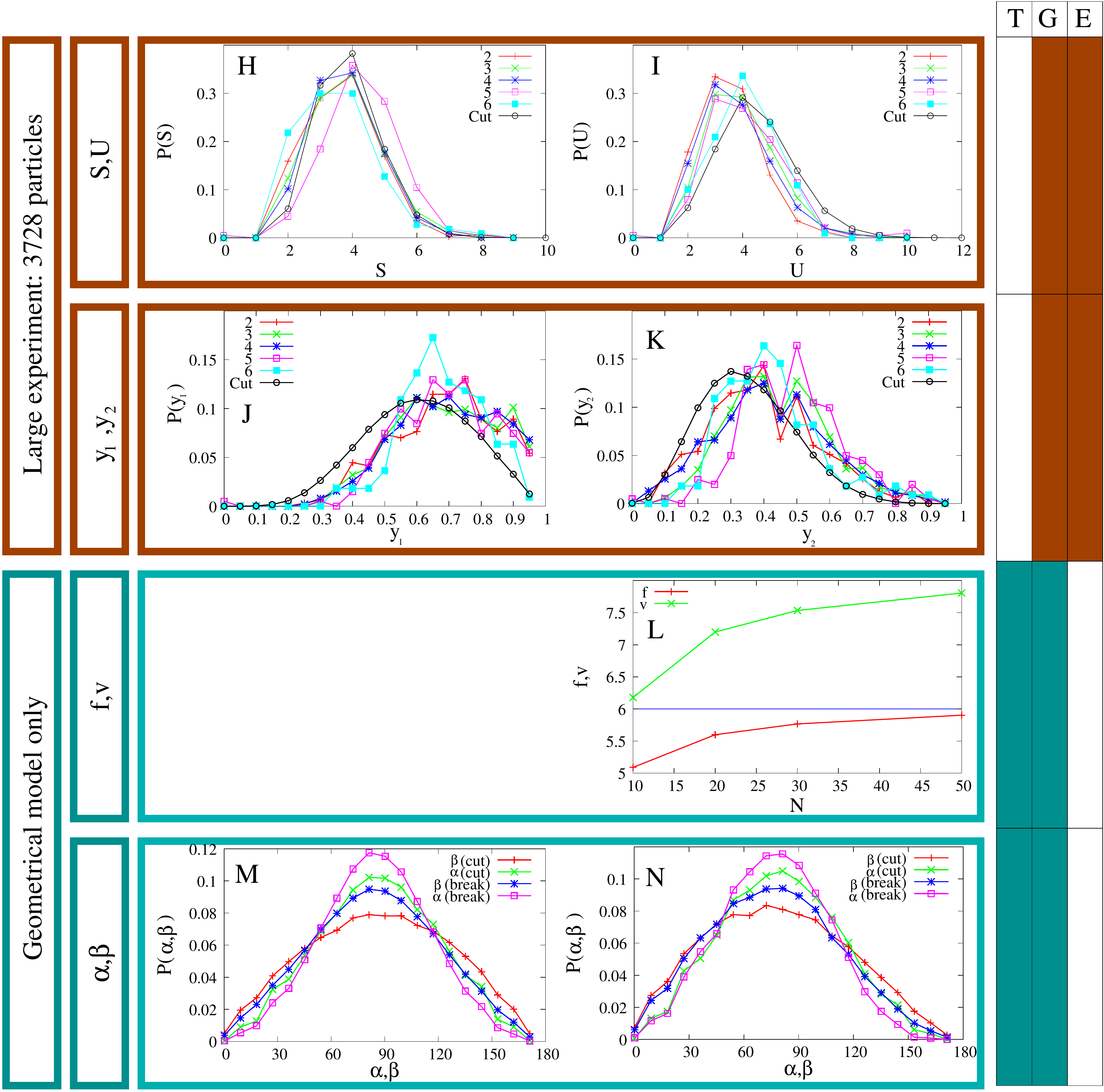}
\caption{\textbf{Fit of the large experimental datasets (Table
\ref{tab:expsource}, 2-6) with the geometric model.} First row (H and I): stable and unstable equilibria ($S,U$), Second row (J and K): axis ratios ($y_1,y_2$).
Fitting parameters for the first and second rows: $[m_0,\sigma_0]=[5\cdot10^{-5},0.05]$ for all data,  with $R_\mathrm{max}^2 \geq 0.95$. The lower two panels show the results of the geometric model. Third row (L): variation of face and vertex-number averages $\bar f,\bar v$ as a function of the number $N$ of applied cutting planes in the cut model. Fourth row, panel M: angle (face and edge) distribution in the fragments of the cut and break models. Panel N: angle (face and edge) distribution in the fragments of the cut and break models, after size selection. }
\label{fig:big2}
\end{center}
\end{figure}

Figures \ref{fig:big1} and \ref{fig:big2} --- the figure was split into two for size reasons --- offer a systematic comparison between
all four listed sources of information. They may be regarded as an extended version of Table \ref{tab:datasource}. In essence, we compared all datasets for all available shape descriptors. The vertical bars on the right summaries the relevant part of Table \ref{tab:datasource}; they indicate which of the three source types (Theory, Geometry and Experiment) was involved in the comparison. As we can observe, we found good match between the most important features. Below we comment on the results illustrated in Figures \ref{fig:big1} and \ref{fig:big2}.

Figure \ref{fig:big1} uses the small dataset with 556 particles, where we had all shape descriptors available and we achieved a match with $R_\mathrm{max}^2 \geq 0.99$  for the illustrated seven density functions by adjusting only the 3 fitting parameters: $[m_0,\sigma_0,A_0]=[6\cdot10^{-4},0.05,0.1]$ for the cut and $[m_0,\sigma_0,A_0]=[10^{-4},0.05,0.1]$ for the break model. The quality of fit is high, especially for the cut model. 

We also mention that we have a good match for cuboid averages in $f,v$ between the field data and theory. The histogram for $v$ shows a striking disparity between odd and even values and this 
is closely matched by the geometric model. As we pointed out earlier, polyhedral cells in primitive mosaics are always \emph{simple} polyhedra, and these always have vertices of third order while the number of these vertices is even. This is well reflected (albeit in an imperfect manner) in the histograms. The imperfection in experimental data is understandable in terms of the fitting parameters. The geometric model matches these imperfections by artificially removing very small faces, which we believe is a fair representation of the limits of measuring the natural fragments by hand; this results in vertices of higher order.
The mass distribution plot is slightly curved on the log-log scale, and is closely matched by the geometric model over more than 1.5 orders of magnitude.

Panels $H,I,J,K$ (rows 1 and 2, brown color) of Fig.~\ref{fig:big2} are related to the large dataset.
Here we could match the $S,U,y_1,y_2$ histograms between experiment and the geometric model, while the theory has no
prediction for \emph{any} of these shape descriptors. Fitting this massive dataset to the computations serves as a validation of the latter, and of the fitting parameters. The relevant fitting parameters were $[m_0,\sigma_0]=[5\cdot10^{-5},0.05]$ for all data, with $R_\mathrm{max}^2 \geq 0.95$.

The fit of the big data shows more variance relative to the small data set, and the fit for aspect ratio is particularly poor. As we pointed out, measurements of $S,U$ were less accurate for this dataset.
Since we tried to minimize error simultaneously for all histograms, this measurement error leads to a worse match on the
$y_1,y_2$ plots.

Panel $L$ (third row) of Fig.~\ref{fig:big2} shows the average number of faces and vertices in the cut model as function of the number $N$ of cutting planes. The observed $\bar f, \bar v$
averages converge to the cuboid averages of $[\bar f,\bar v]=[6,8]$ predicted by theory.

Panels $M$ and $N$ (fourth row) of Fig.~\ref{fig:big2} show face and edge angle distributions observed in the cut and break models. Panel $M$ shows the distribution for all fragments while panel $N$ shows only those that are larger than $m_0=10^{-4}$. The theory of convex mosaics predicts cuboid averages (90 degrees) for primitive mosaics for all angles, and indeed we can observe a marked maximum at that value in the histograms. We remark that the break model shows a sharper peak.
If we select only larger fragments (panel $N$) then the peak shifts to the left, indicating smaller angles.

\clearpage

\bibliography{pnas}

\begin{thebibliography}{10}

\bibitem{turcotte_fractals_1997}
DL Turcotte, {\em Fractals and chaos in geology and geophysics}.
\newblock (Cambridge University Press), (1997).

\bibitem{kawamura_revmodphys_2012}
H Kawamura, T Hatano, N Kato, S Biswas, BK Chakrabarti, Statistical physics of
  fracture, friction, and earthquakes.
\newblock {\em\protect\JournalTitle{Rev. Mod. Phys.}} \textbf{84}, 839--884
  (2012).

\bibitem{adler_fracture_network_book}
PM Adler, JF Thovert, {\em Fractures and Fracture Networks}.
\newblock (Springer, Dordrecht), (1999).

\bibitem{prasher1987crushing}
CL Prasher, {\em Crushing and grinding process handbook}.
\newblock (John Wiley \& Sons), (1987).

\bibitem{mallard_subduction_2016}
C Mallard, N Coltice, M Seton, RD Müller, PJ Tackley, Subduction controls the
  distribution and fragmentation of {Earth}’s tectonic plates.
\newblock {\em\protect\JournalTitle{Nature}} \textbf{535}, 140--143 (2016).

\bibitem{mcewen_europa_1986}
AS McEwen, Tidal reorientation and the fracturing of jupiter's moon europa.
\newblock {\em\protect\JournalTitle{Nature}} \textbf{321}, 49--51 (1986).

\bibitem{trowbridge2016vigorous}
A Trowbridge, H Melosh, J Steckloff, A Freed, Vigorous convection as the
  explanation for pluto’s polygonal terrain.
\newblock {\em\protect\JournalTitle{Nature}} \textbf{534}, 79 (2016).

\bibitem{cuzzi_saturn_2010}
JN Cuzzi, et~al., An evolving view of saturn{\textquoteright}s dynamic rings.
\newblock {\em\protect\JournalTitle{Science}} \textbf{327}, 1470--1475 (2010).

\bibitem{brooker_mars_2018}
L Brooker, et~al., Clastic polygonal networks around lyot crater, mars:
  Possible formation mechanisms from morphometric analysis.
\newblock {\em\protect\JournalTitle{Icarus}} \textbf{302}, 386 -- 406 (2018).

\bibitem{Brilliantov9536}
N Brilliantov, et~al., Size distribution of particles in
  saturn{\textquoteright}s rings from aggregation and fragmentation.
\newblock {\em\protect\JournalTitle{Proceedings of the National Academy of
  Sciences}} \textbf{112}, 9536--9541 (2015).

\bibitem{domokos_plato_2019}
G Domokos, Z L\'angi, Plato's error and a mean field formula for convex
  mosaics.
\newblock {\em\protect\JournalTitle{Axiomathes}} (2019).

\bibitem{Plato_Timaeus}
Plato, {\em Timaeus (translated by B. Jowett)}.
\newblock (Aeterna Press. Original edition MacMillan and Co. London, 1892.),
  (2015).

\bibitem{grunbaum1994uniform}
B Gr{\"u}nbaum, Uniform tilings of 3-space.
\newblock {\em\protect\JournalTitle{Geombinatorics}} \textbf{4}, 49--56 (1994).

\bibitem{ishii_fragmentation_1992}
T Ishii, M Matsushita, Fragmentation of long thin glass rods.
\newblock {\em\protect\JournalTitle{J. Phys. Soc. Jpn.}} \textbf{61},
  3474--3477 (1992).

\bibitem{oddershede_self-organized_1993}
L Oddershede, P Dimon, J Bohr, Self-organized criticality in fragmenting.
\newblock {\em\protect\JournalTitle{Phys. Rev. Lett.}} \textbf{71}, 3107
  (1993).

\bibitem{wittel_fragmentation_2004}
FK Wittel, F Kun, HJ Herrmann, BH Kr\"oplin, Fragmentation of shells.
\newblock {\em\protect\JournalTitle{Phys. Rev. Lett.}} \textbf{93}, 035504
  (2004).

\bibitem{timar_new_2010}
G Tim\'ar, J Bl\"omer, F Kun, HJ Herrmann, New universality class for the
  fragmentation of plastic materials.
\newblock {\em\protect\JournalTitle{Phys. Rev. Lett.}} \textbf{104}, 095502
  (2010).

\bibitem{kun_scaling_2006}
F Kun, FK Wittel, HJ Herrmann, BH Kr\"oplin, KJ Maloy, Scaling behaviour of
  fragment shapes.
\newblock {\em\protect\JournalTitle{Phys. Rev. Lett.}} \textbf{96}, 025504
  (2006).

\bibitem{domokos_universality_2015}
G Domokos, F Kun, AA Sipos, T Szab\'o, Universality of fragment shapes.
\newblock {\em\protect\JournalTitle{Scientific Reports}} \textbf{5}, 9147
  (2015).

\bibitem{ma_shape_2018}
G Ma, W Zhou, Y Zhang, Q Wang, X Chang, Fractal behavior and shape
  characteristics of fragments produced by the impact of quasi-brittle spheres.
\newblock {\em\protect\JournalTitle{Powder Technology}} \textbf{325}, 498 --
  509 (2018).

\bibitem{szabo_reconstructing_2015}
T Szabó, G Domokos, JP Grotzinger, DJ Jerolmack, Reconstructing the transport
  history of pebbles on {Mars}.
\newblock {\em\protect\JournalTitle{Nature Communications}} \textbf{6}, 8366
  (2015).

\bibitem{steacy_automaton_1991}
S Steacy, C Sammis, An automaton for fractal patterns of fragmentation.
\newblock {\em\protect\JournalTitle{Nature}} \textbf{353}, 250--252 (1991).

\bibitem{astrom_universal_2004}
JA Astr\"om, F Ouchterlony, RP Linna, J Timonen, Universal dynamic
  fragmentation in d dimensions.
\newblock {\em\protect\JournalTitle{Phys. Rev. Lett.}} \textbf{92}, 245506
  (2004).

\bibitem{dershowitz1988characterizing}
W Dershowitz, H Einstein, Characterizing rock joint geometry with joint system
  models.
\newblock {\em\protect\JournalTitle{Rock mechanics and rock engineering}}
  \textbf{21}, 21--51 (1988).

\bibitem{goering_evolv_pattern_2013}
L Goehring, Evolving fracture patterns: columnar joints, mud cracks and
  polygonal terrain.
\newblock {\em\protect\JournalTitle{Philosophical Transactions of the Royal
  Society A: Mathematical, Physical and Engineering Sciences}} \textbf{371},
  20120353 (2013).

\bibitem{ma2019universal}
X Ma, J Lowensohn, JC Burton, Universal scaling of polygonal desiccation crack
  patterns.
\newblock {\em\protect\JournalTitle{Physical Review E}} \textbf{99}, 012802
  (2019).

\bibitem{desiccation_book_nakahara}
L Goehring, A Nakahara, T Dutta, S Kitsunezaki, S Tarafdar, {\em Desiccation
  Cracks and their Patterns: Formation and Modelling in Science and Nature}.
\newblock (John Wiley \& Sons), (2015).

\bibitem{goering_softmat_2010}
L Goehring, R Conroy, A Akhter, WJ Clegg, AF Routh, Evolution of mud-crack
  patterns during repeated drying cycles.
\newblock {\em\protect\JournalTitle{Soft Matter}} \textbf{6}, 3562--3567
  (2010).

\bibitem{columnar_joint_science_1988}
A Aydin, JM Degraff, Evoluton of polygonal fracture patterns in lava flows.
\newblock {\em\protect\JournalTitle{Science}} \textbf{239}, 471--476 (1988).

\bibitem{columnar_joint_jagla_pre2002}
EA Jagla, AG Rojo, Sequential fragmentation: The origin of columnar
  quasihexagonal patterns.
\newblock {\em\protect\JournalTitle{Phys. Rev. E}} \textbf{65}, 026203 (2002).

\bibitem{hofmann_prl_2015}
M Hofmann, R Anderssohn, HA Bahr, HJ Wei\ss{}, J Nellesen, Why hexagonal basalt
  columns?
\newblock {\em\protect\JournalTitle{Phys. Rev. Lett.}} \textbf{115}, 154301
  (2015).

\bibitem{cho2019crack}
HJJ Cho, NB Lu, MP Howard, RA Adams, SS Datta, Crack formation and self-closing
  in shrinkable, granular packings.
\newblock {\em\protect\JournalTitle{Soft matter}} (2019).

\bibitem{clair2015geophysical}
JS Clair, et~al., Geophysical imaging reveals topographic stress control of
  bedrock weathering.
\newblock {\em\protect\JournalTitle{Science}} \textbf{350}, 534--538 (2015).

\bibitem{voigtlander2019breaking}
A Voigtl{\"a}nder, M Krautblatter, Breaking rocks made easy. blending stress
  control concepts to advance geomorphology.
\newblock {\em\protect\JournalTitle{Earth Surface Processes and Landforms}}
  \textbf{44}, 381--388 (2019).

\bibitem{molnar2007tectonics}
P Molnar, RS Anderson, SP Anderson, Tectonics, fracturing of rock, and erosion.
\newblock {\em\protect\JournalTitle{Journal of Geophysical Research: Earth
  Surface}} \textbf{112} (2007).

\bibitem{dibiase2018fracture}
RA DiBiase, MW Rossi, AB Neely, Fracture density and grain size controls on the
  relief structure of bedrock landscapes.
\newblock {\em\protect\JournalTitle{Geology}} \textbf{46}, 399--402 (2018).

\bibitem{sklar2017problem}
LS Sklar, et~al., The problem of predicting the size distribution of sediment
  supplied by hillslopes to rivers.
\newblock {\em\protect\JournalTitle{Geomorphology}} \textbf{277}, 31--49
  (2017).

\bibitem{tensilerock_book_2005}
D Bahat, A Rabinovitch, V Frid, {\em Tensile Fracturing in Rocks}.
\newblock (Springer-Verlag Berlin Heidelberg), (2005).

\bibitem{schneider2008stochastic}
R Schneider, W Weil, {\em Stochastic and integral geometry}.
\newblock (Springer Science \& Business Media), (2008).

\bibitem{mills1991fractography}
K Mills, {\em Fractography}.
\newblock (American Society of Metals (ASM) handbook, volume 12), (1991).

\bibitem{bahat2005fractography}
D Bahat, V Rabinovitch, A. adn~Frid, {\em Tensile Fracturing in Rocks}.
\newblock (Springer, Berlin), (2005).

\bibitem{domokos2019honeycomb}
G Domokos, Z L{\'a}ngi, On some average properties of convex mosaics.
\newblock {\em\protect\JournalTitle{Experimental Mathematics}} (2019).

\bibitem{bohn2005hierarchical}
S Bohn, L Pauchard, Y Couder, Hierarchical crack pattern as formed by
  successive domain divisions.
\newblock {\em\protect\JournalTitle{Physical Review E}} \textbf{71}, 046214
  (2005).

\bibitem{anderson2002many}
DL Anderson, How many plates?
\newblock {\em\protect\JournalTitle{Geology}} \textbf{30}, 411--414 (2002).

\bibitem{bird2003updated}
P Bird, An updated digital model of plate boundaries.
\newblock {\em\protect\JournalTitle{Geochemistry, Geophysics, Geosystems}}
  \textbf{4} (2003).

\bibitem{hernandez_physa_1995}
G Hern\'andez, HJ Herrmann, Discrete models for two- and three-dimensional
  fragmentation.
\newblock {\em\protect\JournalTitle{Physica A: Statistical Mechanics and its
  Applications}} \textbf{215}, 420 -- 430 (1995).

\bibitem{kekalainen_solution_2007}
P Kekalainen, JA Astr\"om, J Timonen, Solution for the fragment-size
  distribution in a crack-branching model of fragmentation.
\newblock {\em\protect\JournalTitle{Phys. Rev. E}} \textbf{76}, 026112 (2007).

\bibitem{lamur_disclosing_2018}
A Lamur, et~al., Disclosing the temperature of columnar jointing in lavas.
\newblock {\em\protect\JournalTitle{Nature Communications}} \textbf{9}, 1432
  (2018).

\bibitem{moeraki_boulders_1985}
JR Boles, CA Landis, P Dale, {The Moeraki Boulders; anatomy of some septarian
  concretions}.
\newblock {\em\protect\JournalTitle{Journal of Sedimentary Research}}
  \textbf{55}, 398--406 (1985).

\bibitem{astin1986septarian}
T Astin, Septarian crack formation in carbonate concretions from shales and
  mudstones.
\newblock {\em\protect\JournalTitle{Clay Minerals}} \textbf{21}, 617--631
  (1986).

\bibitem{flinn1962diagram}
D Flinn, On folding during three dimensional progressive deformation.
\newblock {\em\protect\JournalTitle{Quart. Journ. Geol. Soc. London}}
  \textbf{118}, 385--428 (1962).

\bibitem{plimpton1995fast}
S Plimpton, Fast parallel algorithms for short-range molecular dynamics.
\newblock {\em\protect\JournalTitle{Journal of computational physics}}
  \textbf{117}, 1--19 (1995).

\bibitem{brendel2011contact}
L Brendel, J T{\"o}r{\"o}k, R Kirsch, U Br{\"o}ckel, A contact model for the
  yielding of caked granular materials.
\newblock {\em\protect\JournalTitle{Granular Matter}} \textbf{13}, 777--786
  (2011).

\bibitem{national1996rock}
NR Council, , et~al., {\em Rock fractures and fluid flow: contemporary
  understanding and applications}.
\newblock (National Academies Press), (1996).

\bibitem{Szabo_universal_2018}
T Nov{\'a}k-Szab{\'o}, et~al., Universal characteristics of particle shape
  evolution by bed-load chipping.
\newblock {\em\protect\JournalTitle{Science Advances}} \textbf{4} (2018).

\bibitem{domokos_natural}
G Domokos, Natural numbers, natural shapes.
\newblock {\em\protect\JournalTitle{Axiomathes}} (2018).

\bibitem{federico1975polygons}
P Federico, Polyhedra with 4 to 8 faces.
\newblock {\em\protect\JournalTitle{Geometriae Dedicata}} \textbf{3}, 469--481
  (1975).

\bibitem{muller2008age}
RD M{\"u}ller, M Sdrolias, C Gaina, WR Roest, Age, spreading rates, and
  spreading asymmetry of the world's ocean crust.
\newblock {\em\protect\JournalTitle{Geochemistry, Geophysics, Geosystems}}
  \textbf{9} (2008).

\end{thebibliography}

\end{document}